\documentclass[usenatbib,useAMS]{mn2e}
\usepackage{graphicx}
\usepackage{ctable}
\usepackage{url}
\usepackage{times}
\usepackage{xspace}

\newcommand{\lsun}{~\mathrm{L_{\odot}}}

\newcommand{\lir}{L_{\rm IR}}

\newcommand{\msun}{~\mathrm{M_{\odot}}}
\newcommand{\msunperyr}{~\mathrm{M_{\odot} {\rm ~yr}^{-1}}}

\newcommand{\mstar}{M_{\star}}
\newcommand{\mdust}{M_\mathrm{dust}}

\newcommand{\ldust}{L_\mathrm{d}}

\newcommand{\av}{A_{\mathrm{V}}}

\newcommand{\sunrise}{\textsc{sunrise}\xspace}

\newcommand{\gadgetthree}{\textsc{gadget-3}\xspace}

\newcommand{\magphys}{\textsc{magphys}\xspace}
\newcommand{\hatlas}{{\it H}-ATLAS}

\newcommand{\twbc}{T_{\rm W}^{\rm BC}}
\newcommand{\tcism}{T_{\rm C}^{\rm ISM}}
\newcommand{\tc}{T_{\mathrm{C}}^{\mathrm{ISM}}}
\newcommand{\tw}{T_{\mathrm{W}}^{\mathrm{BC}}}
\newcommand{\fmusfh}{f_{\mu}^{\rm SFH}}

\newcommand{\tauv}{\hat{\tau}_{\mathrm{V}}}
\newcommand{\tauvism}{\hat{\tau}_{\mathrm{V,ISM}}}

\newcommand{\pchisq}{(a)\xspace}
\newcommand{\pmstar}{(c)\xspace}
\newcommand{\pldust}{(d)\xspace}
\newcommand{\ptc}{(k)\xspace}
\newcommand{\ptw}{(l)\xspace}
\newcommand{\pmdust}{(e)\xspace}
\newcommand{\pssfr}{(f)\xspace}
\newcommand{\psfr}{(g)\xspace}
\newcommand{\pfmu}{(h)\xspace}
\newcommand{\ptau}{(i)\xspace}
\newcommand{\ptauism}{(j)\xspace}
\newcommand{\pav}{(b)\xspace}

\newcommand{\pchisqn}{a\xspace}
\newcommand{\pmstarn}{c\xspace}
\newcommand{\pldustn}{d\xspace}
\newcommand{\ptcn}{k\xspace}
\newcommand{\ptwn}{l\xspace}
\newcommand{\pmdustn}{e\xspace}
\newcommand{\pssfrn}{f\xspace}
\newcommand{\psfrn}{g\xspace}
\newcommand{\pfmun}{h\xspace}
\newcommand{\ptaun}{i\xspace}
\newcommand{\ptauismn}{j\xspace}
\newcommand{\pavn}{b\xspace}

\newcommand{\fiducial}{{\sf fiducial}\xspace}
\newcommand{\smc}{{\sf SMC-dust}\xspace}
\newcommand{\lmc}{{\sf LMC-dust}\xspace}
\newcommand{\agnoff}{{\sf AGN-off}\xspace}
\newcommand{\altism}{{\sf alternate-ISM}\xspace}

\newcommand{\tmer}{t - t({\rm SFR_{max}})}

\newcommand{\acknowledgments}{\begin{small}\section*{Acknowledgments}\end{small}}

\title[Should we believe UV-mm SED modelling?]{Should we believe the results of UV-mm galaxy SED modelling?}
\author[C.~C. Hayward and D.~J.~B. Smith]{
\parbox[t]{\textwidth}{
Christopher C. Hayward$^{1,2}$\thanks{E-mail: cchayward@caltech.edu}\thanks{Moore Prize Postdoctoral Scholar in Theoretical Astrophysics}
and Daniel J.~B.~Smith$^3$
}
\vspace*{6pt} \\
$^1$TAPIR 350-17, California Institute of Technology, 1200 E. California Boulevard, Pasadena, CA 91125, USA \\
$^2$Heidelberger Institut f\"ur Theoretische Studien, Schloss--Wolfsbrunnenweg 35, 69118 Heidelberg, Germany \\
$^3$Centre for Astrophysics, Science \& Technology Research Institute, University of Hertfordshire, Hatfield, Herts, AL10 9AB, UK
}

\begin{document}

\date{Submitted to MNRAS}

\pagerange{1--24} \pubyear{2014}

\maketitle

\label{firstpage}

\begin{abstract}
Galaxy spectral energy distribution (SED) modelling is a powerful tool, but constraining how well it is able to infer the true values for galaxy
properties (e.g. the star formation rate, SFR) is difficult because independent determinations are often not available. However, galaxy simulations can provide
a means of testing SED modelling techniques. Here, we present a numerical experiment in which we apply the SED
modelling code \magphys to ultraviolet (UV)--millimetre (mm) synthetic photometry generated from hydrodynamical simulations of an isolated disc galaxy and a major galaxy
merger by performing three-dimensional dust radiative transfer. We compare the properties inferred from the SED modelling with the true
values and find that \magphys recovers
most physical parameters of the simulated galaxies well. In particular, it recovers consistent parameters irrespective of the viewing angle, with smoothly varying results for
neighbouring time steps of the simulation, even though each viewing angle and time step is modelled independently.
The notable exception to this rule occurs when we use an SMC-type intrinsic dust extinction curve
in the radiative transfer calculations. In this case, the two-component dust model used by \magphys is unable to effectively correct for the attenuation of the
simulated galaxies, which leads to potentially significant errors (although we obtain only marginally acceptable fits in this case).
Overall, our results give confidence in the ability of SED modelling to infer physical properties of galaxies, albeit with some caveats.
\end{abstract}


\section{Introduction} \label{S:intro}

A galaxy's spectral energy distribution (SED) encodes much information about the galaxy, including its star formation history (SFH);
its stellar, gas, and metal content, and the physical conditions of its interstellar medium (ISM). The number of galaxies, both local and high-redshift,
with well-sampled ultraviolet (UV) to millimetre (mm) integrated SEDs has increased rapidly in recent years, and much effort is being made to attempt to
extract galaxy properties from these SEDs. Accurately extracting physical properties, such as stellar mass and star formation rate (SFR),
from galaxy SEDs is crucial to answer many open questions in galaxy formation, including the following: what is the SFH of the universe?
How do properties such as the SFR, metallicity, and gas fraction depend on redshift and galaxy mass? What processes
quench star formation in galaxies? Furthermore, knowledge of galaxies' physical properties is often necessary to compare observations with theoretical models
because models typically do not directly predict observables.

The simplest method to determine some property of a galaxy is to use a single photometric data point. For example, if the redshift is known, SFRs are
commonly derived from UV, H$\alpha$, or 24-\micron~photometry \citep{Kennicutt:1998review}, and stellar mass can be derived from a galaxy's
near-infrared (NIR) flux \citep[e.g.][]{Bell:2001}. However, such methods require various simplifying assumptions and can suffer from significant systematics
and degeneracies.

Use of multiple data points simultaneously can yield more information and break some degeneracies, such as that between age and metallicity.
In a technique known as SED modelling or stellar population synthesis\footnote{In this work, we use the more-general term `SED modelling' rather than
`stellar population synthesis' because the former can include additional sources of radiation, such as active galactic nuclei (AGN) and dust, beyond direct
stellar emission.}
(e.g. \citealt{Leitherer:1999,Bolzonella:2000}; \citealt[][hereafter BC03]{Bruzual:2003ck}; \citealt{LeBorgne:2004};
\citealt{Maraston:2005}; \citealt{Burgarella:2005,Cunha:2008cy,Kotulla:2009,Kriek:2009,Noll:2009,Conroy:2010b,Serra:2011};
see \citealt{Walcher:2011} and \citealt{Conroy:2013} for reviews), a galaxy is treated as the sum of its parts:
as input, one must use template SEDs for single-age stellar populations (SSPs), which depend on the age and metallicity of the stellar population,
the stellar initial mass function (IMF), and the stellar libraries used.
By assuming an SFH and metallicity, which may be a function of age, the total SED of the stellar population can be calculated. In addition to
stellar emission, nebular emission lines \citep[e.g.][]{Charlot:2001} and AGN emission \citep[e.g.][]{Burgarella:2005,Noll:2009,Berta:2013} can also be included. Dust attenuation
can be treated using an empirical attenuation curve \citep[e.g.][]{Calzetti:1994im,Calzetti:2000iy,Calzetti:1997hu} or a simple analytic model
\citep*[e.g.][hereafter CF00]{Charlot:2000bd}.
A large set of templates is generated by varying the model parameters, and, in principle, the parameters of the template SED that best fits the
UV--NIR photometry can be used to infer physical properties of the galaxy, including SFR, age, metallicity, and stellar mass.

The above discussion has largely ignored infrared (IR) emission from dust, but this, too, can be used to infer galaxy properties.
Indeed, if IR data are unavailable, it is possible to mistake a heavily obscured, rapidly star-forming galaxy for a passive galaxy \citep[e.g.][]{Carter:2009}.
The simplest way to interpret a galaxy's far-IR (FIR) through mm emission\footnote{For simplicity and in accordance with convention, throughout
this work, we will use the terms `IR' or `MIR-mm' to denote the wavelength range 8-1000 \micron, which accounts for the bulk of the dust emission.}
is to fit one or more modified blackbodies to the SED, and in doing so
infer the IR luminosity (which can be used to estimate the SFR), effective dust temperature(s), and dust mass (see sections 4.2 and 4.3 of \citealt*{Casey2014}).
IR SED models, such as those of \citet{Dale:2002} or \citet{Draine:2007sings}, can also be used.
Because the UV--NIR and MIR--mm regions of an SED yield complementary information,
it is preferable to use both simultaneously. An example of a simple method to do this is to infer the SFR by using a combination of the UV or H$\alpha$ luminosity
and the IR luminosity to account for both unobscured and obscured star formation (e.g. \citealt{Kennicutt:2007,Kennicutt:2009iv};
 \citealt*{Relano:2009}; \citealt{Wuyts:2011a,Wuyts:2011b,Reddy:2012, Lanz:2013}).

A more sophisticated approach is to use all available data when fitting SEDs. One method is to perform radiative transfer calculations
assuming some simple galaxy geometry \citep[e.g.][]{Silva:1998ju,Efstathiou:2000,Granato:2000,Popescu:2000,
Tuffs:2004,Siebenmorgen:2007,Groves:2008,Michalowski:2010masses,Michalowski:2010production}.
Alternatively, one can take a more empirical approach and treat the FIR SED as a sum of modified blackbodies \citep[e.g.][hereafter dC08]{Cunha:2008cy}
or use IR SED templates \citep[e.g.][]{Noll:2009}.
Regardless of the manner in which the IR SED is obtained, the luminosity of the IR SED should be equal to the luminosity absorbed by the dust.
This requirement is necessary for the SED model to be self-consistent and can also enable the model to be more constraining because it
explicitly links the UV--NIR and MIR--mm regions of the SED.

The SED modelling methods described above are very flexible, and they can be applied to large samples of galaxies
\citep[e.g.][]{Kauffmann:2003,Brinchmann:2004,Gallazzi:2005,Salim:2007,Cunha:2010by,Smith:2012}.
However, there are multiple simplifying assumptions and uncertainties inherent in the models \citep[e.g.][]{Conroy:2009ks,Conroy:2010a,Conroy:2010b}, some of which
we will discuss here.
The SEDs depend strongly on the IMF, but the shape of the IMF and whether it is universal are both actively debated (e.g. \citealt{Dave:2008};
\citealt*{vanDokkum:2008,vanDokkum:2010,vanDokkum:2011,Conroy:2012,Hopkins:2013IMF,Narayanan:2012IMF,Narayanan:2013};
\citealt{Hayward:2013number_counts}; see \citealt{Bastian:2010vo} for a review).
The SSP templates are also uncertain; one particular area of disagreement is the treatment of thermally pulsating asymptotic giant branch (TP-AGB) stars
\citep{Maraston:2005,Maraston:2006fn,Kriek:2010wr,Henriques:2011,Zibetti:2013}. SED modelling codes typically assume relatively simple SFHs, and changing the assumed
form can significantly affect the results of the modelling \citep[e.g.][]{Michalowski:2012,Michalowski2014}.
Furthermore, the models must necessarily assume simple geometries. In many cases, the dust is treated as a foreground screen or mixed slab.
Some models include somewhat more complicated geometries in that they allow the young stars and older stars to be attenuated by different amounts
(e.g. \citealt{Silva:1998ju}; CF00; dC08; \citealt{Groves:2008}), but this geometry is still only a crude approximation to reality.
Spatial variation in metallicity is typically not accounted for (except perhaps indirectly in the form of an age dependence), and uncertain dust composition can significantly
affect the amount and wavelength dependence of the attenuation, the shape of the dust emission SED, and the inferred dust mass. Finally, treating the FIR emission
as one or more modified blackbodies may be problematic if one's goal is to infer physical quantities rather
than simply describe the SED \citep[e.g.][]{Shetty:2009a,Shetty:2009b,Hayward:2012smg_bimodality,Kelly:2012,Smith:2013},
but more sophisticated models, such as those of \citet{Dale:2002} and \citet{Draine:2007sings}, may be able to yield physical insight.

Given the complexity of and assumptions inherent in SED modelling, it is desirable to compare the results of different methods and, if possible, to test
how well the methods can recover the galaxy properties that they intend to recover.
There have been extensive efforts to `internally validate' SED modelling methods, i.e. search for systematics and uncertainties that are inherent in the methods
using either a sample of synthetic SEDs constructed using the same assumptions that are inherent in the SED modelling codes
or samples of real galaxies (dC08; \citealt{Walcher:2008,
Giovannoli:2011,Boquien:2012,Buat:2012,Buat2014,Smith:2012}; see sections 4.4.4 and 4.5.3 of \citealt{Walcher:2011} for an
extensive discussion). `External validation' of the quantities recovered by SED modelling is possible for only a few quantities, such as the SFR and mass-to-light ratio; however,
when there is a discrepancy between e.g. the SFR inferred from the FIR luminosity and SED modelling, it is not clear \textit{a priori} which SFR value is more
accurate \citep[e.g.][]{Hayward2014,Utomo2014}.

Fortunately, it is possible to test some (but definitely not all) of the assumptions inherent in SED modelling by applying SED modelling to synthetic SEDs generated
from semi-analytical models (e.g. \citealt{Lee:2009}; \citealt{Trager:2009}; \citealt*{Pforr:2012,Pforr:2013}; \citealt{Mitchell2013})
or simulations (e.g. \citealt{Wuyts:2009kr,Lanz2014,Michalowski2014}; Torrey et al., submitted) as a type of
controlled numerical experiment. \citet{Wuyts:2009kr} were the first to perform such tests using hydrodynamical simulations.
They were able to investigate discrepancies caused by mismatches between the true SFH in the simulations and that assumed,
different amounts of attenuation for stars of different ages and AGN, metallicity variations, and AGN contamination unaccounted for in the SED modelling.
However, their method for calculating the photometry was relatively simple: first, it is not clear that the \citet{Calzetti:2000iy} attenuation curve, which is
an empirically derived attenuation curve that is meant to be applied to integrated galaxy SEDs of starburst galaxies, should be applied to attenuate
individual lines-of-sight within galaxies. Furthermore, because \citeauthor{Wuyts:2009kr} did not perform radiative transfer, they could not investigate the
effects of scattering. Additionally, they did not account for the obscuration of young stellar clusters on sub-resolution scales. Finally,
because \citeauthor{Wuyts:2009kr} did not calculate dust re-emission, they restricted their SED modelling to synthetic optical--NIR photometry. As we explain
below, we avoid these limitations by performing dust radiative transfer on hydrodynamical simulations.

Now that radiative transfer is routinely applied to three-dimensional (3-D) hydrodynamical simulations (e.g. \citealt{Jonsson:2006}; \citealt*{Jonsson:2010sunrise};
\citealt{Wuyts:2009b,Wuyts:2010,Bush:2010,Narayanan:2010dog,Narayanan:2010smg,Hayward:2011smg_selection,Hayward:2012smg_bimodality,
Hayward:2013number_counts,Hayward2014,Snyder:2011,Snyder:2013})
and increasingly sophisticated SED modelling is applied to observed galaxies (e.g. dC08; \citealt{Cunha:2010by,daCunha:2010b,Buat:2012,Smith:2012,Lanz:2013}),
it is appropriate to revisit the work of \citet{Wuyts:2009kr}; we do so here by performing radiative transfer on hydrodynamical simulations
of a disc galaxy and a galaxy major merger and applying the SED modelling method of dC08, \magphys, to the synthetic photometry. Because \magphys is now very
commonly used \citep[e.g.][]{Cunha:2010by,daCunha:2010b,Wijesinghe:2011,Rowlands:2012,Rowlands2014,Rowlands2014b,Smith:2012,Banerji:2013,Berta:2013,Fu:2013,
Gruppioni:2013,Ivison:2013,Lanz:2013,Bitsakis:2014,Delvecchio:2014,Presotto:2014,Toft:2014} this work should be of great relevance to many researchers.

This approach is critical for our ability to interpret the results of SED fitting. In addition to the aforementioned, our approach has several further advantages over previous attempts
to validate SED modelling methods. For example, because each line of sight to a galaxy is uniquely affected by dust, it is possible that our ability to infer its properties is viewing-angle dependent.
It is difficult to address this issue using real galaxies, although dC08 attempted to do so statistically by using the ratio of the projected major and minor axes as a crude
proxy for viewing angle; this test revealed no evidence for bias in the averaged values of different \magphys parameters across a sample of 1658
Infrared Astronomical Satellite \citep[IRAS;][]{Neugebauer:1984}-selected galaxies. In contrast, our
approach to generating emergent SEDs at different viewing angles for the same temporal snapshot enables us to address this issue directly. Second, it is highly likely that the extent
of our ability to infer the properties of a galaxy depends on the evolutionary stage of that galaxy (e.g. the length of time since the most recent burst of star formation); this
method of external validation enables us to quantify this effect over timescales in excess of a gigayear, a task that is clearly impossible using real galaxies.

One concern with this method of validation is whether the simulations resemble real galaxies well enough that the test is relevant. This concern is one
reason that we utilise idealised simulations in which the progenitor galaxies are constructed `by hand', as our goal is not to form galaxies \emph{ab initio} but
rather to perform a controlled numerical experiment on simulated galaxies with reasonable properties. This approach helps to ensure that the properties
of the simulated galaxies (e.g. the radial and vertical profiles of the discs, gas fraction, and metallicity) are similar to those of real galaxies.

Furthermore, other works that used the same hydrodynamical and radiative transfer codes and similar initial conditions have demonstrated that the SEDs
agree with those of real galaxies: \citet{Jonsson:2010sunrise} found that for a variety of colour-colour plots spanning the UV through submm, the simulated discs
typically occupied regions that are also occupied by real galaxies from the SIRTF Nearby Galaxies Survey \citep[SINGS;][]{Kennicutt:2003,Dale:2007} sample (but the
full variation in real galaxies' colours was not captured by the simulations, likely because of the limited parameter space spanned by the simulations and because no early-type
or interacting galaxies were simulated). \citet{Lanz2014} used a library of $\sim$12 000 synthetic SEDs of simulated isolated and interacting disc galaxies to fit
the UV--FIR SEDs of a subset of isolated and interacting galaxies \citep[originally presented in][]{Lanz:2013} from the \emph{Spitzer} Interacting Galaxies Survey
(N. Brassington et al., in preparation). They found that most of the real galaxy SEDs were reasonably well-fit by one or more of the simulated galaxy SEDs.
\footnote{\citeauthor{Lanz2014} compared some of the physical properties inferred from the observed SEDs using \magphys with the properties of the corresponding
best-fitting simulated SEDs. However, they did not directly apply \magphys to the simulated SEDs and thus only validated \magphys in an indirect manner. Instead,
the focus of \citet{Lanz2014} was a comparison of observed interacting galaxy SEDs with SEDs predicted from simulations.
Here, we present a more direct and more detailed investigation of the ability of \magphys to recover the properties of simulated galaxy SEDs.}
Similarly, the simulated high-redshift galaxy SEDs of \citet{Hayward:2011smg_selection,Hayward:2012smg_bimodality,Hayward:2013number_counts}
provide acceptable fits to the SEDs of 24-\micron-selected starbursts and AGN (Roebuck et al., in preparation). Thus, we are confident that the simulations
are sufficiently reasonable for the purposes of this work.

The remainder of this paper is organised as follows: in Section \ref{S:methods}, we describe the combination of hydrodynamical simulations and dust radiative transfer used
to create the synthetic photometry and the SED modelling code of dC08, \magphys. Section \ref{S:example_fit} presents an example \magphys fit to a synthetic SED.
Sections \ref{S:isolated_disc} and \ref{S:merger} discuss the results of applying \magphys to the SEDs calculated for the isolated disc and galaxy merger simulations,
respectively, using the default \sunrise parameters. Section \ref{S:systematic_uncertainties} investigates the influence of potential sources of systematic error,
such as the treatment of dust attenuation, in the SED modelling procedure. In Section \ref{S:discussion}, we discuss some implications of our results.
Section \ref{S:conclusions} presents our conclusions.

\section{Methods} \label{S:methods}

To investigate the effectiveness of SED modelling, we first generated synthetic UV--mm SEDs by performing dust radiative transfer on hydrodynamical
simulations of an isolated disc galaxy and a major galaxy merger. We then applied \magphys to the synthetic photometry and compared the
physical parameter values inferred by \magphys with the true values for the simulated galaxies.
Note that the comparison was performed in a blind fashion; CCH generated the synthetic photometry and provided it to DJBS without the
corresponding physical parameter values. Then, DJBS fit the synthetic photometry and provided the inferred parameter values to CCH for comparison.
No modifications to the simulations or SED modelling procedure were made after this comparison was performed, and each snapshot was modelled
independently (i.e. \magphys did not `know' that different viewing angles correspond to the same galaxy or that successive snapshots are in any way related). 

Now, we present the key details of our method for calculating SEDs from hydrodynamical simulations and the SED modelling code \magphys.

\subsection{Calculating SEDs of simulated galaxies} \label{S:sims}

This work uses a combination of 3-D \gadgetthree \citep{Springel:2001gadget,Springel:2005gadget} smoothed-particle
hydrodynamics\footnote{Recently, various authors \citep[e.g.][]{Agertz:2007,Springel:2010arepo,Bauer:2012,Keres:2012,Sijacki:2012,Vogelsberger:2012}
have highlighted issues with the standard formulation
of SPH that may cause the results of simulations performed using the technique to be inaccurate. However, for the type of idealised simulations
used in this work, the standard form of SPH yields results that are very similar to those of the more-accurate moving-mesh hydrodynamics
technique \citep{Hayward2014arepo}.}
galaxy simulations and the {\sc Sunrise}\footnote{\sunrise is publicly available at \\ \url{http://code.google.com/p/ sunrise/}.}
\citep{Jonsson:2006sunrise,Jonsson:2010sunrise} Monte Carlo dust radiative transfer code to calculate synthetic SEDs of the simulated galaxies.
The methods have been described in detail elsewhere \citep[e.g.][]{Jonsson:2006,Jonsson:2010sunrise,Hayward:2011smg_selection,
Hayward:2012smg_bimodality}, so we only summarise them briefly here.

The \gadgetthree simulations include star formation following a volume-density-dependent Kennicutt-Schmidt relation \citep{Schmidt:1959,Kennicutt:1998} with
a low-density cutoff, a sub-resolution prescription for the multiphase ISM  \citep[which implicitly includes supernova feedback;][]{Springel:2003},
and a model for black hole accretion and thermal AGN feedback \citep{Springel:2005feedback}. The current work utilises two \gadgetthree simulations.
One is a simulation of an isolated disc galaxy, the \textsf{vc3} model of \citet{Cox:2006}. The initial conditions consist of a dark matter halo and a rotationally supported
exponential disc of gas and stars. The dark matter halo has a \citet{Hernquist:1990} profile with an effective concentration of 9, spin parameter $\lambda = 0.033$,
and circular velocity $V_{200} = 160$ km s$^{-1}$. The exponential disc has a radial scale length of 3.9 kpc, a total mass of $5.6 \times 10^{10} \msun$, and
an initial gas fraction of 40 per cent.  The second simulation is the \textsf{vc3vc3e} model of \citet{Cox:2006}, which is a merger of two of the previously described disc
galaxies. For this simulation, the two disc galaxies were initialised on a parabolic orbit with a pericentric passage distance of 5 kpc and an initial separation of 100 kpc.
The two discs were initially oriented such that their spin axes are specified by the spherical coordinates $(\theta, \phi) = (30^{\circ}, 60^{\circ})$ and $(-30^{\circ},45^{\circ})$
(the `e' orbit of \citealt{Cox:2006}). The masses and gravitational softening lengths for the baryonic (dark matter) particles are $3.9 \times 10^5 \msun$ and 100 pc
($7.6 \times 10^6 \msun$ and 200 pc), respectively. Please see \citet{Cox:2006} for further details of the \gadgetthree simulations.

At 10-Myr intervals, we saved snapshots of the \gadgetthree simulations and processed them with \sunrise, which calculates the emission from the star and black hole
particles present in the \gadgetthree simulations, propagates the emission through the dusty ISM, and calculates the IR re-emission from dust.
The default \sunrise assumptions and parameters used in this work are identical to those used by \citet{Jonsson:2010sunrise}, except that we include
AGN emission as first introduced in \citet{Younger:2009}.

Star particles with ages $>10$ Myr were assigned {\sc starburst99} \citep[SB99;][]{Leitherer:1999,Vazquez:2005} SSP SED templates according to their
ages and metallicities. Younger star particles were assigned templates from \citeauthor{Groves:2008} (\citeyear{Groves:2008}; see also \citealt{Dopita:2005,
Dopita:2006b,Dopita:2006c}), which include emission from the HII and photodissociation
regions that surround young star clusters. The black hole particles were assigned luminosity-dependent templates from \citet{Hopkins:2007}, which are based on
observations of un-reddened quasars. Because the luminosity of the black hole particle(s) is determined self-consistently from the accretion rate in the
\gadgetthree simulations, the AGN contribution varies significantly with time; see Section \ref{S:AGN} for details.
The dust density distribution was
calculated by projecting the \gadgetthree metal density onto a 3-D adaptive mesh refinement grid and assuming a dust-to-metal density ratio of 0.4
\citep{Dwek:1998,James:2002}. \sunrise calculates dust absorption and scattering using a Monte Carlo method.
Our default dust model is the Milky Way (MW) $R_V=3.1$ model of \citet{Weingartner:2001} as updated by \citet{Draine:2007kk}.

The energy absorbed by the dust
is re-emitted in the IR. \sunrise calculates the emission assuming the dust is in thermal equilibrium (except for half of the PAHs with grain size $<100$ \AA;
see \citealt{Jonsson:2010sunrise} for details). To do so, the code determines the thermal equilibrium dust temperature for each grid cell
and grain species by solving the following equation \citep[e.g.][]{Misselt:2001,Jonsson:2010GPU}:
\begin{equation} \label{eq:dust_T}
\int \sigma_j(\lambda) B(\lambda,T_{ij}) d\lambda = \int I_i(\lambda) \sigma_j(\lambda) d\lambda,
\end{equation}
where $\sigma_j$ is the dust absorption cross section for grain species\footnote{A grain `species' refers to grains of a single size and composition.}
$j$, $I_i(\lambda)$ is the local radiation field intensity in the $i$th grid cell\footnote{The
local radiation field includes contributions both from the attenuated emission from stars and AGN and IR emission that is re-radiated by dust; the latter
source can be significant in e.g. the nuclear regions of starbursts, in which the optical depths can be extremely high.},
$T_{ij}$ is the equilibrium temperature of grain species $j$ in the $i$th grid cell\footnote{Note that in principle, in a given radiative transfer calculation, there can be
$i \times j$ distinct dust temperatures. In the simulations presented in this work, $i$ can be as large as $\sim 10^6$ and $j = 220$; thus, the total number
of distinct dust temperatures in a single radiative transfer calculation can be $\sim 10^8$.}, and
$B(\lambda,T_{ij})$ is the Planck function. Because $I_i(\lambda)$ includes a contribution from dust emission, equation (\ref{eq:dust_T}) must be solved iteratively.
Once the equilibrium dust temperatures are determined, the total SED emitted by dust in grid cell $i$ is calculated using
\begin{equation} \label{eq:dust_SED}
L_{\lambda,i} = 4 \pi \sum_j \sigma_j(\lambda) B(\lambda,T_{ij}),
\end{equation}
and a final radiative transfer step is performed to calculate spatially resolved dust emission SEDs for each viewing angle.

\ctable[
	caption = {Viewing angles (i.e. camera positions) used in the \sunrise radiative transfer calculations.\label{tab:cameras}},
	center,
	notespar,
	doinside=\small,
]{ccc}{
	\tnote[a]{Number used to identify viewing angles in Figs. \ref{fig:disc_attenuation} and \ref{fig:merger_attenuation}.}
	\tnote[b,c]{Camera positions ($\theta$ and $\phi$ denote the polar and azimuthal angles, respectively) in spherical coordinates.
	The isolated disc galaxy and merger orbit lie in the $xy$-plane. Thus, angle 1 provides a face-on view of the disc galaxy.}
}{
														\FL
Angle number\tmark[a]	&	$\theta$\tmark[b]	&	$\phi$\tmark[c]	\NN
					&	(deg)				&	(deg)			\ML
1					&	0				&	0			\NN
2					&	73.4				&	0 			\NN
3					&	73.4				&	120			\NN
4					&	73.4				&	240			\NN
5					&	124.8			&	0			\NN
6					&	124.8			&	120			\NN
7					&	124.8			&	240			\LL
}

The results of the \sunrise calculations are spatially resolved UV--mm SEDs (i.e. integral field unit spectrograph-like data) for the simulated galaxies
viewed from seven different cameras. To sample uniformly in solid angle, the positions were selected by uniformly sampling the cosine of the polar angle, $\cos \theta$,
starting at the north pole and excluding the south pole ($\cos \theta = \{-1/3, 1/3, 1\}$). For each $\cos \theta$ value except for $\cos \theta = 1$, for which all azimuthal
angles are equivalent, the azimuthal angle $\phi$ was sampled uniformly ($\phi = \{0, 2\pi/3, 4\pi/3\}$).
The camera positions in spherical coordinates are specified in Table \ref{tab:cameras}.

We calculated the integrated photometry by
summing the SEDs of all pixels and convolving with the appropriate filter response curves. We assumed that the simulated galaxies are at redshift $z = 0.1$.
In this work, we used the bands that were used for the initial
\textit{Herschel} ATLAS \citep[\textit{H}-ATLAS;][]{Eales:2010} investigations because one of the motivations of the work was to validate the SED modelling approach used in
\citet{Smith:2012} for 250\,$\mu$m-selected galaxies with $r$-band Sloan Digital Sky Survey \citep[SDSS;][]{York:2000} counterparts from \citet{Smith:2011}. The
likelihood-ratio cross-matching in \citet{Smith:2011} was performed in order to associate redshift information primarily from SDSS and the Galaxy and Mass Assembly \citep[GAMA;][]{Driver:2011} 
survey with the \textit{H}-ATLAS sources, and to leverage the matched multi-wavelength photometry from GAMA for the purposes of fitting SEDs \citep[we refer the
interested reader to][for further details]{Driver:2011}. Specifically, we used simulated photometry in the near- and far-UV bands of the Galaxy Evolution Explorer satellite
\citep[GALEX;][]{Martin:2005}, the \textit{ugriz} bands from the SDSS,  
the \textit{YJHK} bands from the UK Infrared Deep Sky Survey \citep[UKIDSS;][]{Lawrence:2007}, and FIR data from IRAS at 60 $\micron$, 
\textit{Herschel} PACS \citep{Poglitsch:2010} at 100 and 160 $\micron$, and \textit{Herschel}
SPIRE \citep[][]{Griffin:2010} at 250, 350, and 500 $\micron$. Note that at all times, the photometry
is integrated over the entire system (i.e. both galaxies in the merger).

Because the goal of this work is to test the efficacy of SED modelling under ideal conditions and investigate
intrinsic systematic uncertainties rather than those that arise from noisy or limited data, we did not add any noise to the photometry. However, for the purposes of
applying \magphys, we assumed uncertainties identical to those assumed in \citet{Smith:2012}, amounting to 0.2\,mag in the FUV and NUV bands, 0.1\,mag in the
\textit{ugrizYJHK} bands, 20 per cent at 60 $\micron$, 10 (20) per cent at 100 (160) $\micron$ and 15 per cent at 250, 350 and 500 $\micron$. Although in
\citet{Smith:2012}, the motivation for these uncertainties was to do with issues regarding absolute calibration uncertainties and hard-to-quantify aperture effects
\citep[e.g.][]{Hill:2011}, they are arbitrary for this investigation (but are similar to the quantifiable model uncertainties in the radiative transfer calculations;
\citealt{Lanz2014}).

\subsection{SED modelling} \label{S:dC08}

As previously discussed, we performed SED modelling using \magphys (dC08), which is now very commonly used for interpreting observed galaxy SEDs
(see example references in Section \ref{S:intro}).
We used the  version described in \citet{Smith:2012}, which was used for the \textit{H}-ATLAS analysis therein; here, we summarise the most relevant details,
and we refer the reader to dC08 for full details of the method.

\magphys fits galaxy SEDs using a Bayesian approach to determine posterior distributions for the fit parameters. In this manner, median-likelihood values
for physical properties of a galaxy, such as the SFR, are inferred.
The emission from stars for a given IMF, SFH, and metallicity is determined using the `CB07' (unpublished) version of the BC03 SSPs.\footnote{The CB07 templates are the default templates used in \magphys. However, it has recently been discovered that the CB07 models over-correct
for the contribution of TP-AGB stars \citep{Zibetti:2013}. For this reason, it is now possible to use the BC03 models in \magphys.}
Dust attenuation
is treated via the method of CF00; in this approach, young stars (with age $< 10^7$ Myr) are more attenuated than older stars to account
for stars being born in dense molecular clouds. All stars are attenuated by an effective optical depth $\hat{\tau}_{\lambda}^{\rm ISM}$,
which is given by equation (4) from dC08:
\begin{equation}
\hat{\tau}_{\lambda}^{\rm{ISM}} = \mu \hat{\tau}_{V} (\lambda/5000~{\rm{\AA}})^{-0.7}.
\end{equation}
The young stars are further attenuated by effective optical depth $\hat{\tau}_{\lambda}^{\rm BC}$,
which is given by equation (3) of dC08:
\begin{equation}
\hat{\tau}_{\lambda}^{\rm{BC}} = (1-\mu) \hat{\tau}_{V} (\lambda/5000~{\rm{\AA}})^{-1.3}.
\end{equation}
In the above equations, $\hat{\tau}_V = \hat{\tau}_{\lambda}^{\rm ISM} + \hat{\tau}_{\lambda}^{\rm BC}$ is the total effective optical depth
to the young stars and $\mu = \hat{\tau}_V^{\rm ISM}/(\hat{\tau}_V^{\rm BC} + \hat{\tau}_V^{\rm ISM})$ is the fraction of the total optical
depth contributed by the `diffuse ISM'. The specific power-law indices adopted in the above equations were motivated by fitting
the CF00 model to observations of local starburst galaxies. Note that assuming that the dust has MW-, LMC- or SMC-type properties, the
the CF00 model can be well-reproduced with a discrete-cloud geometry (see CF00 for full details).

The FIR dust emission is treated as a sum of multiple optically thin\footnote{The assumption of optical thinness in the FIR is likely to be reasonable for all but the most
extreme local galaxies, and modelling normal galaxies was the original purpose for which \magphys was designed. However, this assumption may be
problematic for extremely IR-luminous, highly obscured galaxies, such as submm galaxies \citep{Hayward:2012smg_bimodality,
Rowlands2014}.}
modified blackbodies with different normalizations, dust temperatures, and dust emissivity indices, $\beta$. We do not discuss the implementation in full detail here,
instead referring the interested reader to dC08. However, we will briefly highlight some salient features of the \magphys\ dust implementation, in which some of the parameters
are fixed based on observational constraints and some are allowed to vary. The FIR emission is dominated
by `warm' and `cold' grains in thermal equilibrium decomposed into the birth cloud and diffuse ISM components of the CF00 model.   
The birth clouds and diffuse ISM both have warm-dust components, which are treated as modified blackbodies with $\beta = 1.5$; the temperature of the warm birth cloud component, $\twbc$,
is allowed to vary within a prior between $30 \le \twbc \le 60$\,K, whereas the ISM warm component has a fixed temperature of 45\,K. 
In contrast, only the diffuse ISM has a cold-dust component, which is represented as a modified blackbody with $\beta = 2.0$ and variable temperature $15 \le \tcism \le 25$\,K.
For the purpose of calculating dust masses,
the dust emissivity is normalised at $\kappa_{850~\mu {\rm m}} = 0.77$ g$^{-1}$ cm$^2$ \citep{Dunne:2000}.

Given the SED components described above, dC08 generated libraries of template SEDs, including a set of 25 000 stellar population models with a wide
variety of SFHs (which have the general form of an exponentially declining component with superimposed bursts), metallicities, and dust attenuation.
dC08 also generated a separate
set of 50 000 dust SED templates with a range of dust temperatures and relative contributions of different dust components (see dC08 for the
details of the sampling and the assumed prior distributions for the model parameters). 

The separate stellar population and dust emission SED templates are combined to yield UV--mm SEDs.
One of the parameters that describes the stellar population template SEDs is the fraction of the absorbed luminosity that is absorbed by the {diffuse ISM
rather than the birth clouds, $\fmusfh$. Similarly, a parameter for the dust emission template SEDs is the fraction of the IR luminosity that is
emitted by dust in the diffuse ISM, $f_{\mu}^{\rm IR}$. When \magphys\ combines the stellar population and dust emission template SEDs, to make the
SEDs self-consistent (i.e. to satisfy the `energy balance' criterion), it requires that
\begin{equation}
f_{\mu}^{\rm SFH} = f_{\mu}^{\rm IR} \pm \delta f_{\mu},
\label{eq:df}
\end{equation}
where $\delta f_{\mu} = 0.15$. Strict equality is not required to account for uncertainties from e.g.
viewing angle, and dC08 found that $\delta f_{\mu} = 0.15$ was sufficient to yield good fits to observed galaxy SEDs.
This condition requires that the UV-NIR and MIR-mm emission are self-consistent; thus, the availability of UV--mm constraints is leveraged more fully
by the fitting procedure than by treating the UV-NIR and MIR-mm components in isolation.

Applying the condition specified in equation (\ref{eq:df}) to all possible combinations of the 25 000 stellar population and 50 000 dust emission template SEDs
yields a library of millions of UV--mm SED templates. \magphys then fits galaxy SEDs in a Bayesian manner
using the $\chi^2$ estimator to determine the goodness-of-fit (see \citealt{Kauffmann:2003} for an early application of such a technique).
That \magphys uses $\chi^2$ for the SED fitting requires that each datum has an associated error
estimate to appear in the denominator of the $\chi^2$ calculation. In the case of real data, this uncertainty can include several different components, such as photon
shot noise, calibration uncertainties, and aperture effects, which are clearly not applicable to our model 
(which is noise-free, precisely calibrated, and includes integrated photometry). As noted in Section \ref{S:sims}, we arbitrarily adopt uncertainties in each band
identical to those used in the \magphys fits for {\it H}-ATLAS galaxies in \citet{Smith:2012}. This is potentially problematic because if the simulated SEDs were
perfectly represented in the \magphys libraries, this would result in extremely small values of best-fit $\chi^2$, a point to which we shall return when discussing
our results below.

\begin{figure*}
\centering
\includegraphics[width=1.95\columnwidth,trim=0cm 0cm 0cm 0.65cm,clip]{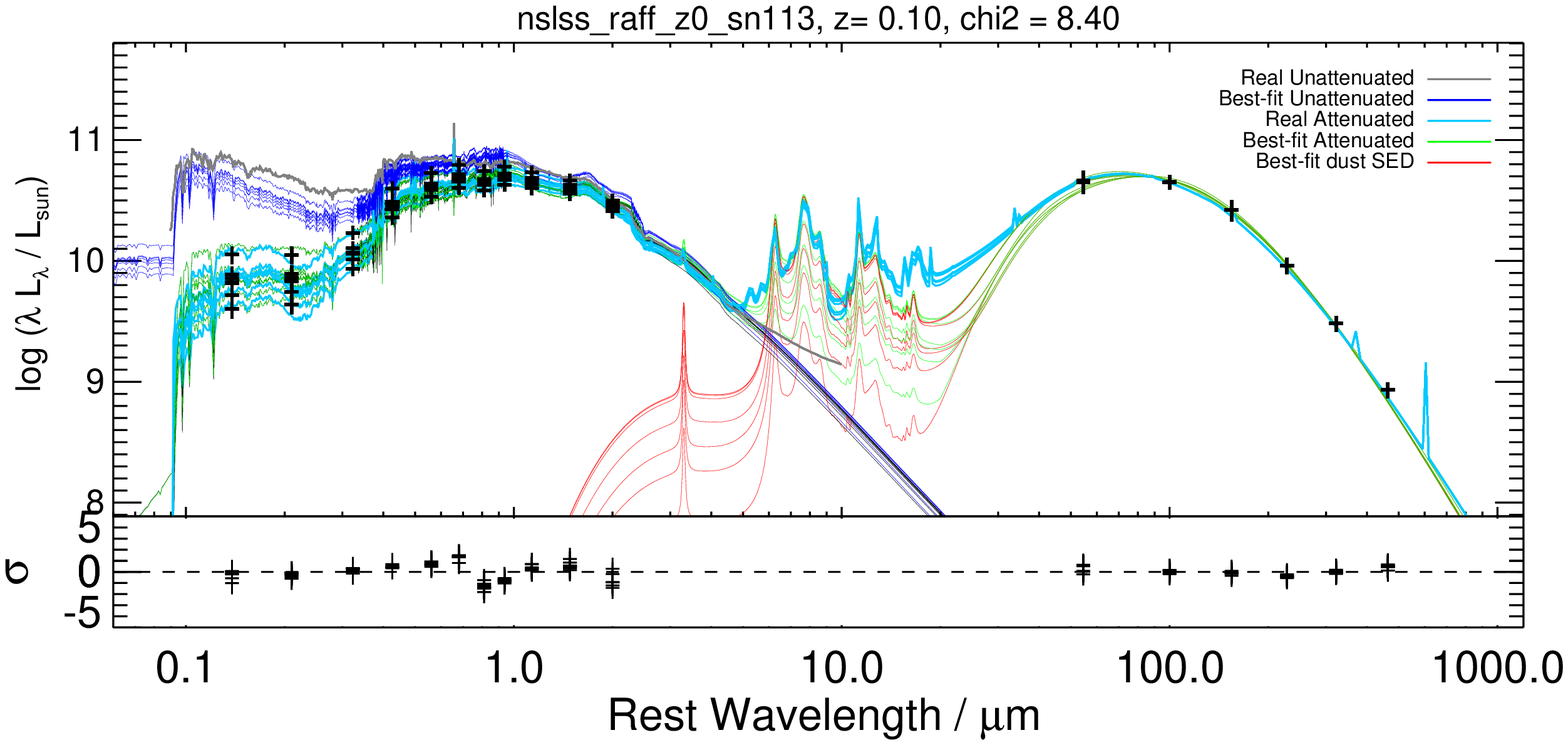}
\caption{The top panel shows an example of the \magphys SED fits for the $\tmer = -0.5$ Gyr snapshot (between first passage and coalescence)
of the merger simulation observed from seven viewing angles. The black points are the simulated photometry. The cyan lines correspond to the
`observed' SEDs of the simulated galaxy for each
of the seven viewing angles, and the grey line indicates the true input (unattenuated) SED. The green, blue, and red lines denote the best-fitting total
output (i.e. the sum of the attenuated emission from stars and the dust emission), unattenuated
stellar, and dust SEDs, respectively, yielded by \magphys for each of the seven viewing angles. The bottom panel shows the residuals in the photometry,
$(L_{\rm true}-L_{\magphys})/\sigma$. For each
of the viewing angles, \magphys yields an acceptable fit to the photometry, and the true intrinsic stellar SED is recovered reasonably well. The larger
residuals near \textit{i} band occur because H$\alpha$ falls in this band at $z = 0.1$, and emission lines are not accounted for in our \magphys modelling;
this has the effect of leaving a positive residual in \textit{i} band and affecting the neighbouring residuals (because neighbouring bands are not independent in SED modelling).
In the MIR, the \magphys SEDs vary strongly with viewing angle, and the true SEDs are not recovered for most of the viewing angles. In the FIR shortward of
the observed-frame 60-$\micron$ data point, \magphys underestimates the true SEDs.}
\label{fig:sed_example}
\end{figure*}

\begin{figure*}
\centering
\includegraphics[width=2.0\columnwidth,trim=0cm 0cm 0cm 0.35cm,clip]{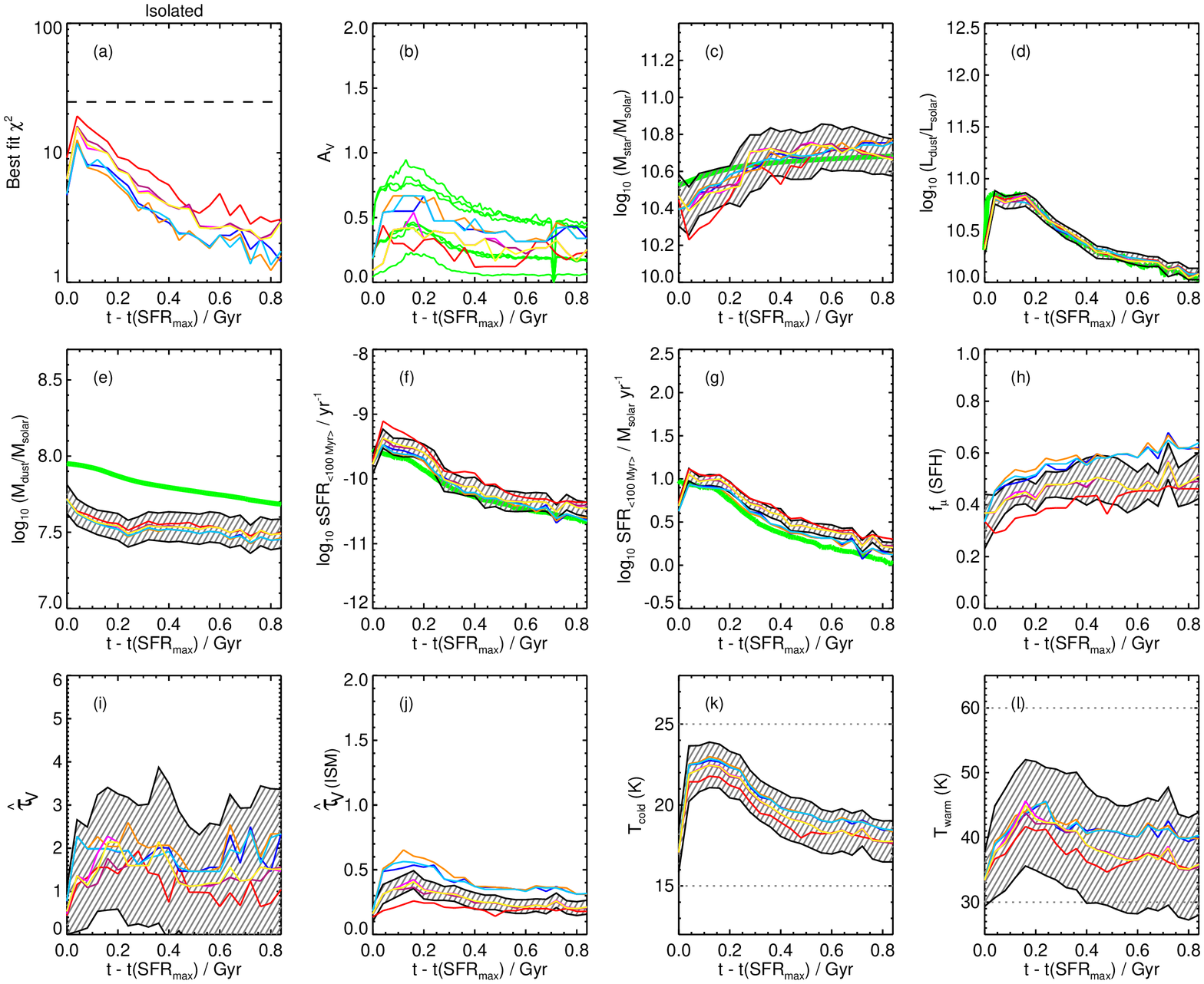}
\caption{Results of applying \magphys to the synthetic integrated photometry for the isolated disc simulation. Each panel shows
the evolution of a \magphys parameter vs. simulation snapshot time $t$ in Gyr. The coloured lines indicate median-likelihood values inferred from \magphys, and different
colours denote different viewing angles  (in order of angle number as specified in Table \ref{tab:cameras}, red, blue, orange, light blue, pink, purple and yellow).
The shaded regions represent the median of the symmetrized 16th and 84th percentiles of the cumulative frequency distribution about the median of each set of parameters
averaged over the seven viewing angles. When possible
(not all \magphys parameters correspond to physical parameters of the simulations), the true values from the simulation are plotted as a solid green line
(in panels \pavn, \pmstarn, \pldustn, \pmdustn, \pssfrn, and \psfrn).
The panels are as follows: \pchisq $\chi^2$ value for the best-fitting SED, where the dashed line indicates the threshold for an acceptable fit from \citet{Smith:2012};
\pav $V$-band attenuation ($\av$);
\pmstar stellar mass; \pldust total luminosity of the dust emission; \pmdust dust mass; \pssfr specific SFR; \psfr SFR;
\pfmu fraction of luminosity absorbed by the diffuse ISM in \magphys ($\fmusfh$); \ptau total $V$-band optical depth in \magphys ($\tauv$);
\ptauism $V$-band optical depth of the diffuse ISM in \magphys ($\tauvism$);
\ptc cold-dust temperature parameter of \magphys ($\tcism$); and \ptw warm-dust temperature parameter of \magphys ($\twbc$) In the last two panels,
the dotted lines represent the limits imposed by the assumed priors.
Most parameters are recovered well; see the text for details.}
\label{fig:disc}
\end{figure*}

\section{Results} \label{S:results}

\subsection{Example fit} \label{S:example_fit}

Fig. \ref{fig:sed_example} shows the results of applying \magphys to the $\tmer = -0.5$ Gyr snapshot of the merger simulation. The spread in the photometric points at a given wavelength
reflects the viewing-angle-dependent variation in dust attenuation, which is self-consistently computed for the simulated galaxy through dust radiative transfer. In the UV,
the output SEDs for the seven cameras span a range of $\sim 0.5$ dex in luminosity over the different viewing angles. For the most-obscured viewing angle, the observed NUV luminosity is an order
of magnitude fainter than the intrinsic luminosity. Longward of $\sim1~\micron$, the variation of the photometry with viewing angle and the attenuation are considerably less (although still non-negligible).

For each of the seven viewing angles, \magphys yields an acceptable fit to the photometry, although the models underpredict the $i$-band data points because at $z = 0.1$
(the assumed redshift of the
simulated galaxy), the high-equivalent-width H$\alpha$ emission line, which is not considered in this implementation of \magphys, falls roughly in the centre of that band's
transmission function.\footnote{Interestingly, this systematic bias is also seen in the {\it H}-ATLAS \magphys\ analysis of 250-\micron-selected galaxies in \citet{Smith:2012}.}
Encouragingly, \magphys is able to recover the intrinsic (unattenuated) stellar SED to within $\sim 0.1-0.4$ dex. This success indicates that the
CF00 two-component dust attenuation model used in \magphys is effective at correcting for the effects of dust attenuation (for this particular SED; we present
an example in which this is not the case in Section \ref{S:dust}). In the simulations, the dust attenuation for a given
viewing angle depends on the 3-D spatial distribution of sources of emission and dust, spatial variations in the stellar populations, and dust scattering into and out of the given
line of sight. Differential extinction is significant because for a given line of sight, there is, in principle, a unique line-of-sight optical depth for each stellar particle; thus, a
two-component model is surely a crude approximation to the actual geometry of the simulated galaxy. Consequently, it is impressive that the dust attenuation correction
is as effective as it is.

\begin{figure}
\centering
\includegraphics[trim=265 0 265 0,clip,width={0.85\columnwidth}]{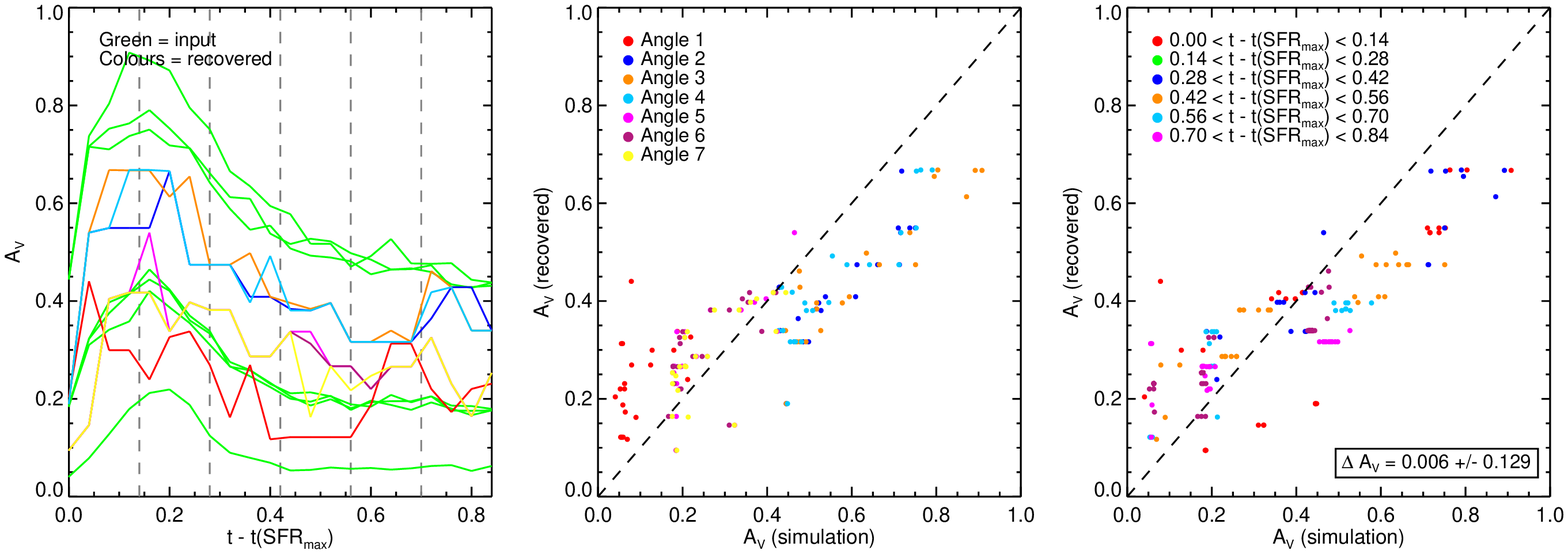}
\caption{$\av$ values of the \magphys best-fitting SEDs vs. the true $\av$ values for the isolated disc simulation. The points are coloured according to the viewing angle, as specified
in the legend. For viewing angles for which the true $\av$ value is relatively low (angles 1, 5, 6, and 7), \magphys tends to overestimate the $\av$ values. Conversely, for viewing angles
with higher $\av$ values (angles 2, 3, and 4), \magphys tends to underestimate $\av$. The average offset between the \magphys and true $\av$ values is $0.006 \pm 0.129$.}
\label{fig:disc_attenuation}
\end{figure}

Using these input data, \magphys is significantly less effective at recovering the true dust SED in the MIR, primarily because of the lack of
`observations' at wavelengths in the range of $\sim2-50~\micron$,
where the MIR SED is poorly recovered for most viewing angles. At these wavelengths, the variation in the best-fitting SEDs for each viewing angle output by \magphys
is greater than an order of magnitude in luminosity (whereas the variation in the true SED is negligible). In the FIR between $\sim 25$ \micron~and the 60-$\micron$ data point,
all of the best-fitting SEDs for this snapshot under-predict the true SED \citep[although this is not necessarily the case for other snapshots, this trend was also noted by][]{Ciesla:2014}.
As noted in \citet{Smith:2012}, these difficulties are not
unexpected because the only constraints on the MIR SED in the absence of MIR observations come from the prior on the MIR component of the dust SED library (which is
chosen at random and thus deliberately broad) and the energy balance criterion. This latter constraint is also weakened in the MIR regime as a result of the small contribution
of the hot dust component to the total dust luminosity. 

The uncertainty in the MIR highlights the fundamentally phenomenological (rather than physical) nature of the model for the IR emission: for a given total energy absorbed, the dust
emission does not depend \textit{a priori} on the SED of the absorbed light. In reality, the shape of the radiation field that heats the dust, which can vary significantly throughout a galaxy,
affects the dust-temperature distribution. For this reason, as noted in dC08 and \citet{Smith:2012}, the efficacy of the dust emission model in \magphys at observationally un-sampled
wavelengths (particularly in the MIR) is limited. The model may be useful for recovering the IR luminosity and dust mass (we address these possibilities below), but it should
not be used to interpret the detailed physical state of the dust or to make predictions for regions of the SED that are unconstrained by the available photometry. Using a more
physically motivated model for dust emission, such as that of \citet{Draine:2007sings}, may alleviate this problem \citep{Ciesla:2014}.

\subsection{Isolated disc} \label{S:isolated_disc}

Fig. \ref{fig:disc} shows the time evolution of various quantities for the isolated disc simulation. In each panel, the thin non-green lines indicate the median-likelihood
values output by \magphys (except for the $\chi^2$ and $\av$ panels, which show the values for the best-fitting SED);
different colours correspond to different viewing angles, and the shaded region
represents an estimate of the typical uncertainty about the median (specifically, it represents the median of the symmetrized 16th and 84th percentiles of the cumulative frequency
distribution about the median of each set of parameters, averaged over the seven viewing angles). The thick green lines represent the true values of the quantities
for the simulations (when possible; not all \magphys parameters have a direct physical counterpart in the simulations). See the figure legend for details of the parameters
shown in each panel.

At all times during this isolated disc simulation, acceptable fits can be found, which is to say that the $\chi^2$ values (shown
in panel \pchisqn) are always below the threshold value for an acceptable fit (shown as the horizontal dashed line). This threshold value of $\chi^2$ was derived
in \citet{Smith:2012} on the basis that it corresponds to the $\chi^2$ value above which there is a probability of less than one percent that the best fit is consistent
with the model given the seventeen bands of photometry available, their total errors, and a statistical estimate of the number of free parameters in the model
(see \citealt{Smith:2012} for the technical details of this derivation). In Section \ref{S:methods}, we have already mentioned the arbitrary nature of the photometric
errors that we have adopted in this study (given the absence of e.g. calibration uncertainties and aperture effects in our simulated photometry). That the best-fit
$\chi^2$ values are non-negligible ($1 < \chi^2 < 20$) highlights that there are differences between the SEDs emergent from the
simulation and the \magphys fitting libraries (which is not surprising because it is unlikely that the simple treatment of dust attenuation used in \magphys can perfectly capture
the relatively complex source and dust geometry of the simulated galaxies). As we shall discuss
below, the generally reasonable parameter estimates that \magphys derives, relative to the known simulated values, suggest that the $\chi^2$ threshold appears
sufficiently large to allow us to confidently recover reasonable SED fits for the simulated SEDs; we shall return to this topic below. 

\begin{figure}
\centering
\includegraphics[width={\columnwidth}]{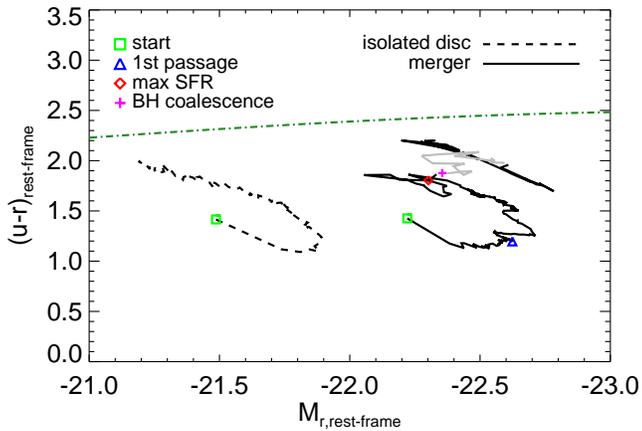}
\caption{SDSS $u-r$ colour vs. $r$-band absolute AB magnitude for the simulated disc galaxy (dashed line) and galaxy merger (solid line). Various times
of interest are marked, as described in the legend. The grey segment of the solid line indicates the time period of the merger simulation during which
\magphys does not yield acceptable fits to the simulated SEDs. The dashed line indicates the optimal separator between the blue cloud and red sequence from
\citet{Baldry:2006}. For most of the duration of both simulations, the simulated galaxies are within the blue cloud. After the final starburst (red diamond),
the simulated merger continues to approach the green valley. Because the merger simulation was terminated $\sim0.5$ Gyr after the starburst, there is not sufficient time for it to
move to the red sequence.}
\label{fig:cmd}
\end{figure}

The physical evolution of the isolated disc is simple:
because there is no gas accretion in this idealised simulation, as time progresses, the gas content is depleted, the SFR decreases, and $\mstar$ increases.
The time evolution of the various simulation quantities is qualitatively recovered by the SED modelling: the physical and fitted values of
$\av$ (panel \pavn), dust luminosity $\ldust$ (panel \pldustn), $\mdust$ (panel \pmdustn), sSFR (panel \pssfrn), and SFR (panel \psfrn) all decrease with time, whereas both the physical and fitted
values of $\mstar$ (panel \pmstarn) increase with time. 

As well as the general trends, it is worth noting that the output parameters vary smoothly within the errors between adjacent time snapshots.
This is reassuring because \magphys\ fits each snapshot (and viewing angle) independently without knowledge that the snapshots\slash angles are related; the lack of discontinuities
in the derived parameters offers considerable support for the reliability of the parameters that \magphys\ produces. 

However, the quantitative agreement between the physical and fitted parameters is more varied. $\ldust$ (panel \pldustn) is recovered exceptionally
well because the simulated photometry samples the FIR SED well, in particular around the peak \citep[e.g.][]{Smith:2013}, and the model SEDs typically provide good fits to the
simulated photometry. Although the inferred and true SFR\footnote{The \magphys SFRs plotted in this work correspond to SFRs averaged over the past
100 Myr, although our results are almost identical if we instead consider \magphys SFRs with 10 Myr averaging. The value for the simulations is the `instantaneous' SFR, i.e.
the sum of the SFRs of the individual gas particles, which are calculated based on their gas
densities and the assumed sub-resolution star formation prescription. Consequently, the SFR value for the simulations corresponds to an average over a shorter
timescale (i.e. less than the maximum time step, 5 Myr) than the \magphys values. If the SFR varies significantly on $10-100$ Myr timescales, this difference
could lead to discrepancies between the \magphys and simulation values even if \magphys recovers the SFH exactly. However, for most times in the simulations,
this effect is minor.} values (panel \psfrn) agree well at early times, the true SFR is increasingly overestimated as the simulation
progresses; the overestimate can be as much as $\sim 0.2$ dex.

\begin{figure*}
\centering
\includegraphics[width=2.0\columnwidth,trim=0cm 0cm 0cm 0.35cm,clip]{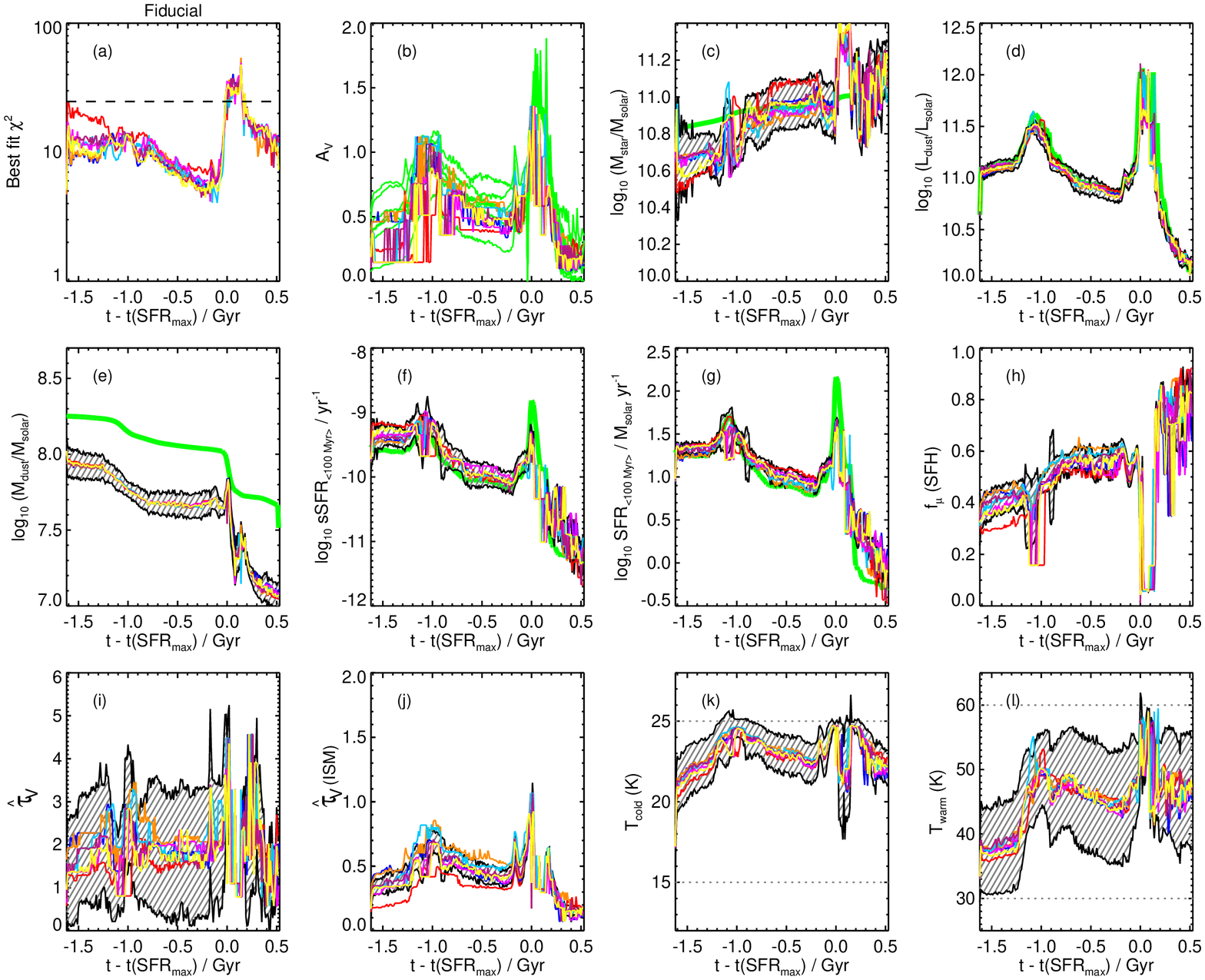}
\caption{Similar to Fig. \ref{fig:disc}, but for the galaxy major merger simulation. The x-axis in each panel indicates the time relative to the peak of the starburst induced at
coalescence. This convention thus provides some insight into the physical state of the system at a given time. The qualitative evolution and values of the various physical parameters are recovered
very well, even during the coalescence-induced starburst phase, when the fits are formally unacceptable. As for the isolated disc case, the dust mass is systematically underestimated.}
\label{fig:merger}
\end{figure*}

The dust mass\footnote{The dust emissivities assumed by the two codes differ: in \magphys, the emissivity is normalised by $\kappa_{850 ~{\rm \mu m}} = 0.77$ g$^{-1}$
cm$^{2}$ \citep{Dunne:2000}, whereas the MW dust model used in the simulations has $\kappa_{850 ~{\rm \mu m}} = 0.38$ g$^{-1}$ cm$^{2}$. Consequently, we multiply the dust
masses output by \magphys by two to account for this difference.} (panel \pmdustn) is systematically underestimated by $\sim 0.2 - 0.3$ dex.
This underestimation is at least partially due to the assumption in the simulations that the cold phase of the sub-resolution ISM has a negligible volume filling factor and
thus does not absorb photons. Consequently, dust contained in the cold phase does not emit light and cannot be recovered. In Section \ref{S:mp_off},
we discuss this issue in more detail.

Within the uncertainties, the inferred and true stellar masses (panel \pmstarn) agree throughout the simulation. However, at early times, the median-likelihood values
can be less than the true values by $\sim0.1-0.3$ dex.
Because the stellar mass is well-recovered and the SFR is slightly overestimated, the specific SFR (panel \pssfrn) is also overestimated slightly.

For some viewing angles, $\av$ (panel \pavn) tends to be underestimated, whereas for others, it is typically overestimated. This is indicated more clearly in
Fig. \ref{fig:disc_attenuation}, which shows the $\av$ recovered by \magphys versus the
true $\av$. For less-attenuated (closer to face-on) viewing angles (angles 1, 5, 6, and 7), $\av$ is slightly overestimated, whereas for
more-attenuated (closer to edge-on) viewing angles, $\av$ is slightly underestimated by \magphys. On average, $\av$ is recovered to within $0.006 \pm 0.129$.
For a given viewing angle, the inferred and true $\av$ values typically differ by less than 0.2 magnitudes.

Although the other plotted quantities do not have direct physical analogues in the simulations, their time evolution is also of interest. Panels \ptc and \ptw show
the \magphys dust temperatures $\tc$ and $\tw$ versus time. $\tc$ and $\tw$ both
tend to decrease as the simulation progresses. This decrease reflects the shifting of the simulated galaxy's SED to longer wavelengths with time because
the strong decrease in luminosity coupled with a relatively weak decrease in the dust mass results in colder dust (see the discussion in \citealt{Hayward:2011smg_selection}).
The median likelihood values for $\tc$ ($\tw$) are in the range $\sim 18-23$ ($33-46$) K, and the uncertainty, which is more significant than the variation
with viewing angle, is $\sim 2$ ($5-10$) K.

The total $V$-band optical depth, $\tauv$ (panel \ptaun), and the $V$-band optical depth contributed by the diffuse ISM, $\tauvism$ (panel \ptauismn),
both remain relatively constant over time.
At all times, the diffuse ISM is optically thin and the total effective optical depth (birth clouds plus diffuse ISM) is $\sim 1-2$, although there is significant variation with both
time and viewing angle, and the uncertainty is relatively large. As the simulation progresses and the sSFR decreases, the fraction of the luminosity absorbed
by the diffuse ISM, $\fmusfh$ (panel \pfmun), increases. The diffuse ISM absorbs of order half of the total luminosity ($\fmusfh \sim 0.3-0.6$).

The variation with viewing angle is indicated by the differences among the \magphys parameter values at a given time. For most parameters, the variation
is less than the \magphys uncertainties\footnote{Note that the none of the viewing angles are edge-on (but three have relatively high inclinations
of $73.4$ deg). Had we used an edge-on camera, the overall variation among
viewing angles would certainly be greater. However, it is unlikely that our conclusions regarding the importance of viewing angle variation would change
qualitatively, and the probability of observing real disc galaxies almost perfectly edge-on is low.}
(i.e. the coloured lines lie within the shaded region). The notable exceptions are $\tauvism$ and, to a lesser extent,
$\fmusfh$. Physically, $\tauvism$ should vary with viewing angle because as the disc is viewed closer to edge-on, the typical column depths along the
line of sight are greater. For the same reason, $\fmusfh$, which is the fraction of the absorbed stellar light that is absorbed by the diffuse ISM rather than
the birth clouds, should also vary with viewing angle. The physical viewing-angle-dependent variation in the obscuration is demonstrated by panel \pav
of Fig. \ref{fig:disc} and further highlighted in Fig. \ref{fig:disc_attenuation}; as discussed above, the true $\av$ values (the green lines) typically differ by only $\sim 0.2$\,mag from the best-fit estimates.

\subsection{Major galaxy merger} \label{S:merger}

We now turn to the evolution of the major galaxy merger simulation. This merger exhibits the characteristic evolution of major mergers that induce strong
starbursts (not all orbits result in such starbursts; see \citealt{Cox:2006}).
Because the progenitor galaxies are not initialised with bulges, which tend to stabilise the discs, a starburst with maximum SFR of $\sim 60 \msunperyr$ is
induced at first pericentric passage ($\tmer \sim -1.1$ Gyr). Subsequently, the SFR decreases below the initial value. As the progenitor disc galaxies
approach final coalescence ($\tmer \sim 0$ Gyr), a starburst that is even stronger than that at first passage is induced by the tidal torques exerted
by the galaxies upon one another. The SFR and $\ldust$ briefly exceed $100 \msunperyr$ and $10^{12} \lsun$, respectively (i.e. the simulated galaxy would be
classified as an ultraluminous IR galaxy, ULIRG). Shortly after the peak of the starburst, the AGN contribution (see Section \ref{S:AGN}) is maximal; the AGN can contribute
as much as 75 per cent of the total UV--mm luminosity \citep[see e.g.][for a detailed study of a real merger-induced starburst in a ULIRG exhibiting AGN activity]{Smith:2010}.
During the final starburst, a significant fraction of the available gas is consumed. Shock heating
and AGN feedback heat the bulk of the remaining gas (see e.g. \citealt{Hayward2014arepo} for details). Consequently, the SFR plummets from $\sim 100 \msunperyr$
to less than $\sim 0.5 \msunperyr$, and the AGN emission decreases rapidly.

To put the simulated merger in context, the time evolution of the simulated merger in the SDSS $u-r$ colour versus $r$-band absolute AB magnitude $M_r$
colour-magnitude diagram (CMD) is shown in Fig. \ref{fig:cmd}. For completeness, the time evolution of the isolated disc is also shown. During most of the duration of the simulations,
the galaxies are in the blue-cloud region of the CMD \citep[see e.g.][]{Baldry:2004,Baldry:2006,Darg:2010}. After the starburst that is induced at final coalescence
of the merging galaxies, the simulated merger continues its evolution towards the green valley, the locus of which is denoted by the green dot-dashed line \citep[from][]{Baldry:2006}.
Because the simulation was terminated $\sim0.5$ Gyr after the peak of the starburst, the system does not transition onto the red sequence.
Our aim is to investigate how well \magphys can fit the SEDs of actively star-forming, IR-luminous galaxies, for which the full panchromatic
capabilities of \magphys can be utilised. Thus, the fact that the simulated merger does not enter the red sequence is irrelevant for the purposes of this work.

It is worthwhile to note that the simulated galaxies occupy regions of the CMD that are populated by real galaxies \citep[e.g.][]{Baldry:2004,Baldry:2006,Darg:2010}.
This is true even in the phase of the merger simulation during which \magphys is unable to yield an acceptable fit to the simulated SEDs; this time period is indicated
by the grey segment of the solid line in Fig. \ref{fig:cmd}.

Fig. \ref{fig:merger} shows the results of applying \magphys to the SEDs of the merger simulation; the panels are the same as in Fig. \ref{fig:disc}.
As was the case for the isolated disc, \magphys qualitatively recovers the true evolution of the physical parameters except
for the stellar mass and dust mass for a short time near merger coalescence. For example, the times and amplitudes of the starbursts
are captured exceptionally well. The success of \magphys at inferring the time evolution of the merger is reassuring and perhaps even surprising because (1)
the version of \magphys used here was designed to treat relatively normal local galaxies, not ULIRGs; (2) \magphys does not include emission from AGN, which is
significant at some times (near coalescence) during this merger simulation (the issue of AGN contamination will be discussed in detail in Section \ref{S:AGN}); 
(3) \magphys treats each viewing angle and
time snapshot individually without knowledge of one another; and (4) because of the first
two reasons, \magphys does not formally achieve a good fit to the SEDs during the coalescence stage of the merger [i.e. for $-0.1 \la \tmer \la 0.2$ Gyr, the
$\chi^2$ value is greater than the threshold for an acceptable fit; see panel {\pchisqn}].

For most of the snapshots and viewing angles, the values inferred by \magphys for most of the parameters are consistent with the true values within the uncertainties.
Because the FIR photometry are typically well-fit by \magphys, $\ldust$ (panel \pldustn) is recovered extremely well. The SFR (panel \psfrn) is also typically
recovered well; note that this is not necessarily a consequence of the excellent recovery of $\ldust$ because \magphys\ includes a possible contribution to the
dust luminosity from evolved stars that are not linked with the most recent burst of star formation.
In the pre-coalescence phase [$-1 \la \tmer \la -0.2$ Gyr], the SFR tends to be overestimated slightly (by $\la 0.1$ dex), and in the post-starburst
phase, it can be overestimated by as much as 0.6 dex (which is a smaller factor than would occur if a simple conversion from $\lir$ were used; see \citealt{Hayward2014}),
irrespective of whether we consider \magphys SFRs averaged over 10 or 100 Myr (i.e. the \magphys default 100 Myr SFR-averaging timescale is not the source of this discrepancy).
The stellar mass (panel \pmstarn) is recovered to within $\sim 0.2$ dex except during the final coalescence\slash starburst phase, when the fits are statistically unacceptable.
At early times, it is systematically underestimated.

In Sections \ref{S:methods} and \ref{S:isolated_disc}, we discussed the choice of $\chi^2$ threshold that we use to identify bad fits, a threshold that is exceeded during
the coalescence phase at the time of the peak starburst and AGN activity. That the threshold is exceeded here offers further encouragement for our arbitrary choice of photometric errors:
using this $\chi^2$ threshold, we are able to get an acceptable fit to $\sim 95$ per cent of the snapshots, and it is only during the $\sim 5$ per cent of the simulation when the
starburst and AGN activity are most intense that we are unable to derive a good fit to the simulated photometry. This time period is when the physics of the galaxy and the \magphys library are
most discrepant, and visual inspection of the `best-fitting' SEDs suggests that these fits should be rejected.\footnote{This effect also raises the tantalising possibility of using poor fits as a means
of identifying sources that have undergone recent mergers, though as discussed in \citet{Smith:2012}, there are several other possible reasons for poor fits (e.g. errors in the photometry, incorrect
cross-identification, and\slash or artificially narrow prior libraries).} To summarize, although we adopt uncertainties on the simulated
data out of necessity for the purposes of applying \magphys rather than because of the physical effects that blight real data, the results that they produce do seem to be at
least plausible and the resulting threshold value appears to function broadly as expected.

Returning to the recovered parameters, the sSFR (panel \pssfrn) is typically recovered within the uncertainties, although at early times,
the median-likelihood values from \magphys can be as much as $\sim 0.5$ dex greater
than the true values because of the underestimate of $\mstar$ at these times.
The sSFR is slightly overestimated in the post-starburst phase because
of the overestimate of the SFR at these times.

The dust mass (panel \pmdustn) is systematically underestimated by $\sim 0.2-0.5$ dex.
However, as for the isolated disc, a significant part of the underestimate is because in the simulation,
by construction, the dust in the `cold phase' of the sub-resolution ISM does not
absorb or emit radiation. We investigate and discuss this issue in detail in
Section \ref{S:mp_off}.

\begin{figure}
\centering
\includegraphics[trim=265 0 265 0,clip,width={0.85\columnwidth}]{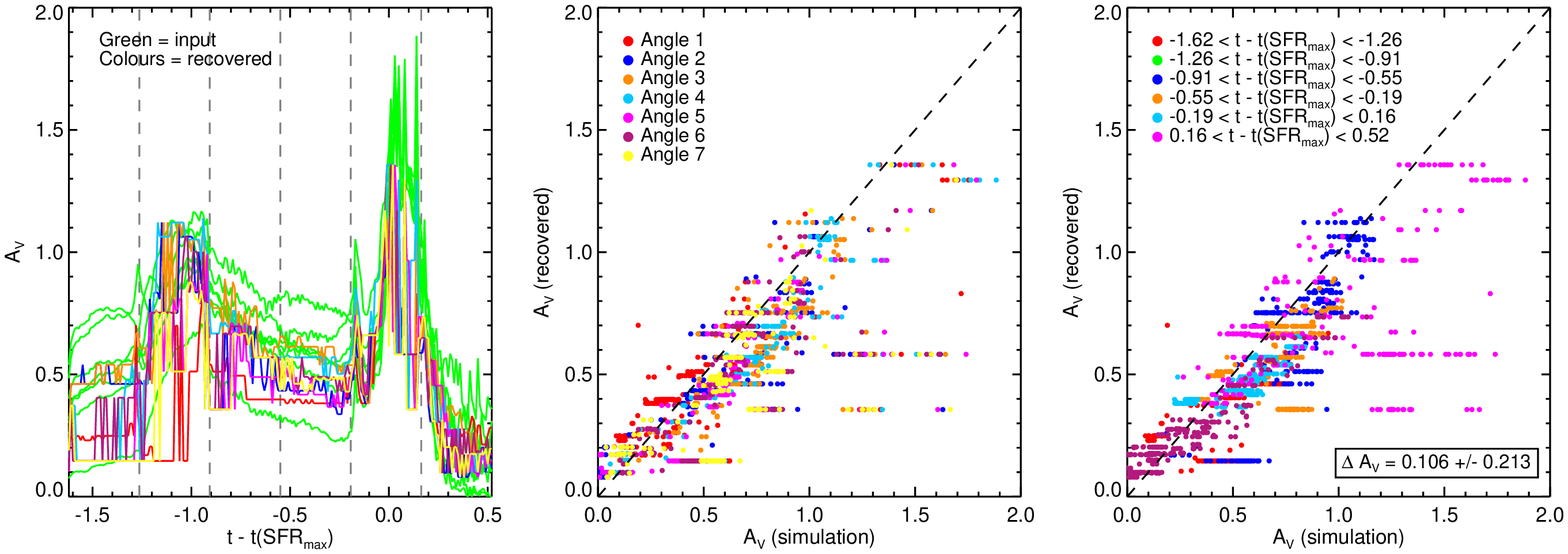} \\
\includegraphics[trim=530 0 0 0,clip,width={0.85\columnwidth}]{./plots/compare_av_fiducial.eps}
\caption{Median-likelihood $\av$ values from \magphys versus true $\av$
for all snapshots of the merger simulation. In the top panel, the points
are coloured according to the viewing angle, whereas in the bottom panel,
the colours indicate the time relative to the peak of the final starburst. For
most of the simulation, \magphys recovers the true $\av$ to within $\sim 0.2$
dex. However, near the time of the final starburst (the cyan points in the lower
panel), $\av$ can be underestimated by as much as $\sim 1$ magnitude.
Unlike for the isolated disc simulation, there is no significant viewing-angle
dependence.}
\label{fig:merger_attenuation}
\end{figure}

The time evolution of $\av$ is shown in panel \pav, but how well it is recovered
can be read more easily from Fig. \ref{fig:merger_attenuation}. For most of the
merger simulation, the true $\av$ is recovered to within $\sim 0.2$ mag. The average
offset $\Delta \av = A_{\mathrm{V, true}} - A_{\mathrm{V, \magphys}} = 0.106 \pm 0.213$.
The upper panel of Fig. \ref{fig:merger_attenuation} indicates that, unlike for the
isolated disc simulation, there is no significant viewing-angle dependence because
there are no `special' viewing angles for this highly asymmetric system.
The lower panel shows that there is no systematic offset between the true and
inferred $\av$ values, except for during the time period near the final starburst (indicated by the light-blue symbols),
when the $\av$ tends to be underestimated by \magphys, sometimes by greater than 1 magnitude; however, as we have previously mentioned, we are unable to
derive acceptable fits to the photometry during this stage of the merger.

The evolution of the \magphys parameters that have no direct physical analogue in the simulations still provides some interesting insights into the physical
state of the simulated galaxies. The cold (panel \ptcn) and warm (panel \ptwn) dust temperatures are in the range $\sim 20-25$ and $\sim 36 - 60$ K, respectively.
Thus, they tend to be higher for the merger than for the isolated disc. The formal uncertainties are similar to those for the isolated disc, $\sim 1-2$ and $\sim 5-10$
K for $\tc$ and $\tw$, respectively. Both temperatures increase sharply during the starburst induced at first pericentric
passage ($\tmer \sim -1.15$ Gyr); this behaviour reflects the increase in effective dust temperature (i.e. the shift of the IR SED
peak to shorter wavelength; e.g. \citealt{Smith:2013}) that is caused in starbursts primarily by the simultaneous sharp increase in luminosity and decrease in dust mass
\citep{Hayward:2011smg_selection}.\footnote{This effect is also seen around the peak associated with the merger coalescence, but the unacceptably high $\chi^2$
values during this stage of the simulation preclude any physical interpretation.} Although it is encouraging that the evolution of the \magphys dust temperatures reflects
this physical effect, this result should be interpreted with caution because of the significant error bars associated with $\tc$ and $\tw$ and the proximity of the
median-likelihood values to the bounds on the temperature priors (dotted grey horizontal lines in panels \ptcn and \ptwn).

The effective optical depths (panels \ptaun and \ptauismn) are especially interesting. For most of the simulation, $\tauvism \la 1$. Interestingly, $\tauvism$ peaks during both starbursts, which is
physically reasonable because the emission at those times is dominated by relatively compact, obscured starbursts. The variation with viewing angle is small, but
it is greater than the formal uncertainty on $\tauvism$. The typical total optical depth $\tauv$ is $\sim 2$, but this value varies significantly with time and viewing angle
and is very uncertain (i.e. at fixed time and viewing angle, the confidence interval can span the range $\tauv \sim 0-4$).

The fraction of the total luminosity absorbed by the diffuse ISM, $\fmusfh$ (panel \pfmun), decreases sharply during the starbursts.
This result is consistent with the physical expectation that in starbursts, the dust luminosity is dominated by highly obscured young stars. This
is certainly true in the simulations, and it is impressive that the \magphys parameter evolution reflects this effect, even when the fits are formally
unacceptable during the starburst induced at merger coalescence. However, the decrease in $\fmusfh$ during the starbursts may simply
be a consequence of the assumed prior because by construction, only the birth clouds contain dust with $T > 45$ K.

As for the isolated disc case, most of the modelled parameters show little or no
viewing-angle dependence once the uncertainties are taken into account. The only exception is $\tauvism$; as
explained above, this quantity should vary with viewing angle, whereas
quantities such as the SFR and stellar mass should not.

\ctable[
	caption = {\sunrise runs used to investigate systematic uncertainties.\label{tab:runs}},
	center,
	star,
	notespar,
	doinside=\small,
]{ll}{
	\tnote[a]{Run designation.}
	\tnote[b]{Description of the assumptions used in the \sunrise calculations.}	
}{
																												\FL
Designation\tmark[a]	&	Description\tmark[b]																				\ML
\fiducial			&	{\sc starburst99} SSP templates, AGN emission enabled, MW-type dust, default (clumpy) sub-resolution ISM model		\NN
\agnoff			& 	AGN emission disabled (solely in the radiative transfer calculations; see footnote \ref{footnote:agn_off})				\NN	
\lmc				& 	LMC-type dust used instead of MW-type dust															\NN
\smc				& 	SMC-type dust used instead of MW-type dust															\NN
\altism			& 	Alternate sub-resolution ISM model (no sub-resolution clumpiness)											\LL
}

\subsection{Systematic uncertainties} \label{S:systematic_uncertainties}

In this section, we present tests in which we performed additional \sunrise radiative transfer calculations on the merger simulation. In each test, we varied
one of the assumptions in the \sunrise calculation and kept all others identical to those of the default-parameter run. The assumptions used for these tests are summarised
in Table \ref{tab:runs}. These tests enable us to characterise how our ignorance of the underlying `microphysics', such as the details of stellar evolution and the dust grain
composition, affects the accuracy of the SED modelling.

\begin{figure}
\centering
\includegraphics[width={0.99\columnwidth}]{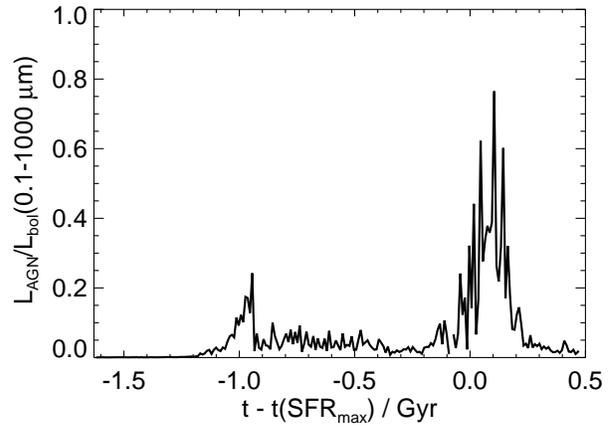} \\
\caption{Fractional contribution of the AGN(s) to the total 1-1000 $\micron$ luminosity versus time. The AGN contribution is most significant during the time
periods shortly after first pericentric passage and final coalescence. The maximum AGN contribution, $\sim 75$ per cent of the 1-1000 $\micron$ luminosity,
is $\sim 100$ Myr after the coalescence-induced starburst.}
\label{fig:agn_frac}
\end{figure}

\begin{figure}
\centering
\includegraphics[trim= 0 377 531 11,clip,width={0.75\columnwidth}]{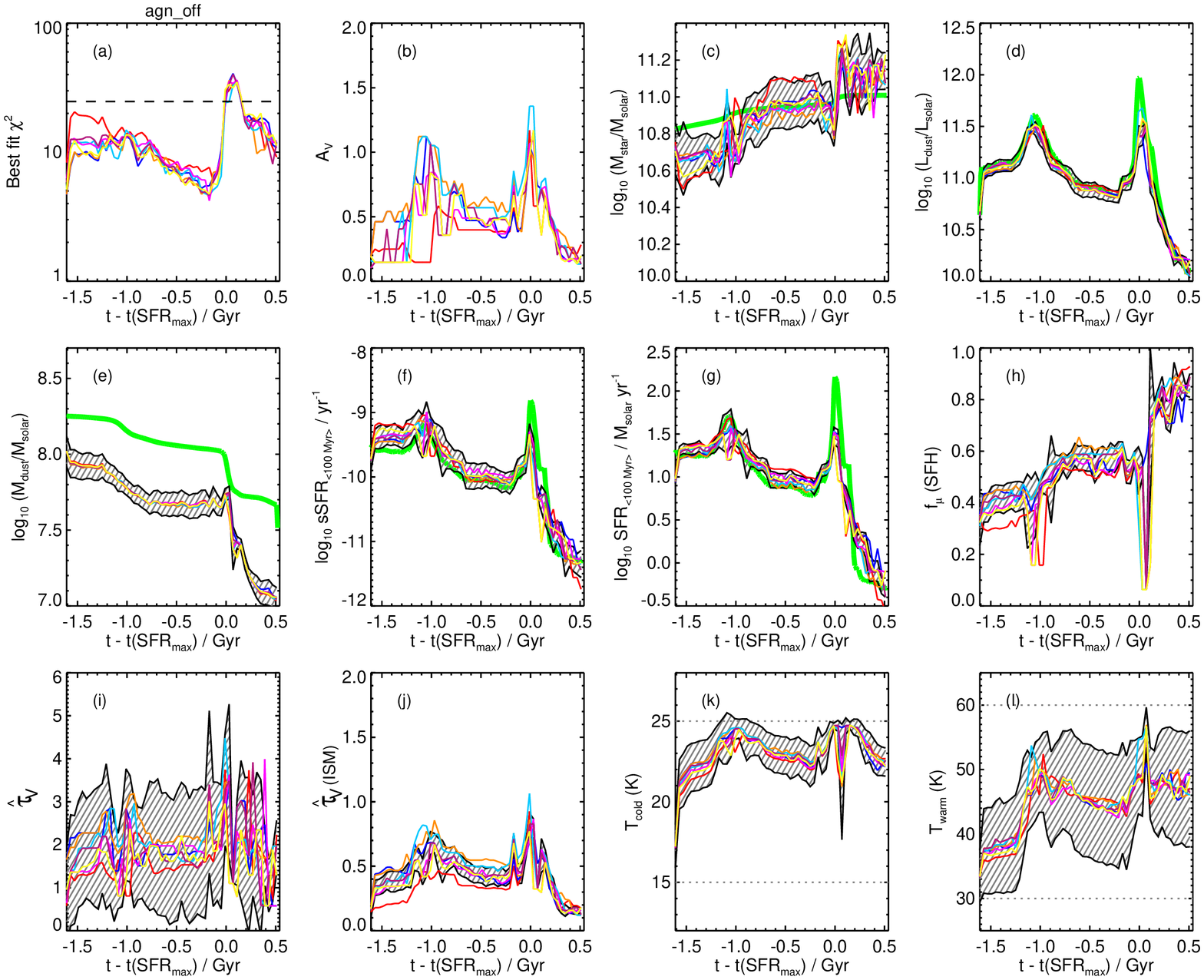}\\
\includegraphics[trim= 354 377 177 0,clip,width={0.75\columnwidth}]{./plots/fitresults_agn_off.eps}
\caption{Selected results for the \agnoff test. At the time of maximum AGN luminosity
[$\tmer \sim 0.2$ Gyr], the $\chi^2$ values for the best-fitting model (top panel) are less than for the \fiducial run, which demonstrates that AGN contamination hinders the ability
of \magphys to obtain a satisfactory fit at the time when the AGN is most active. Note that the stellar mass (bottom panel) is recovered more accurately
than for the \fiducial run, which indicates that AGN contamination partially causes the overestimate of the stellar mass at that time in the \fiducial run.}
\label{fig:agn_off}
\end{figure}

\begin{figure}
\centering
\includegraphics[trim=0 377 531 11,clip,width={0.75\columnwidth}]{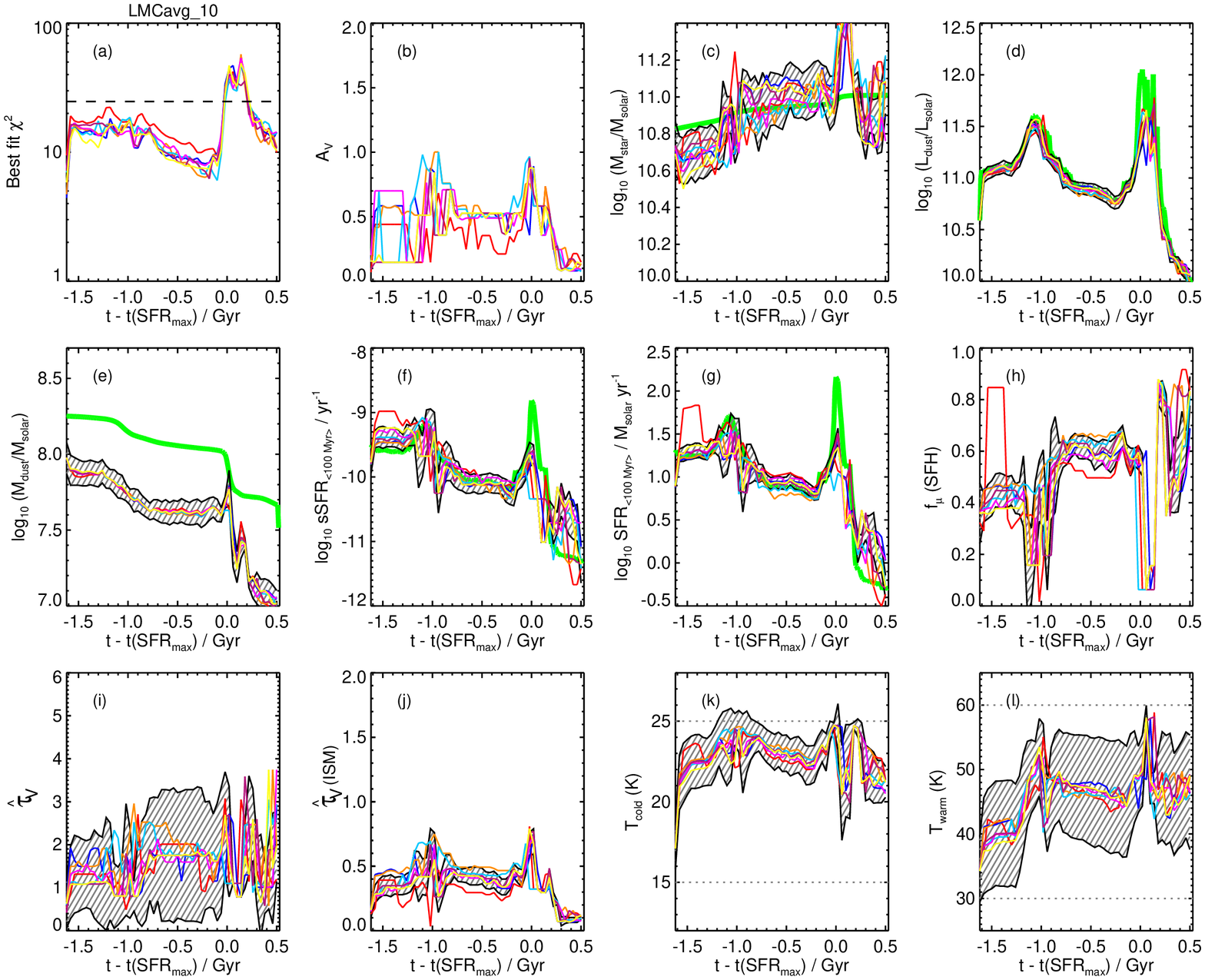} \\
\includegraphics[trim=354 377 177 0,clip,width={0.75\columnwidth}]{./plots/fitresults_lmcavg_10.eps}
\caption{Selected results for the \lmc test.
When LMC-type dust is used to calculate mock SEDs of the simulated merger, the $\chi^2$ values (top panel) and stellar mass (bottom panel) differ considerably from the \fiducial
case; for all other parameters, the time evolution does not differ significantly from the \fiducial case. For the \lmc run, the $\chi^2$ values are slightly higher. The
median-likelihood stellar mass values are systematically $\sim 0.1$ dex greater, but for most of the evolution of the merger, they are still consistent with the true values.}
\label{fig:lmc}
\end{figure}

\subsubsection{AGN contamination} \label{S:AGN}

The \fiducial \sunrise runs include emission from the AGN particles in the \gadgetthree simulation. The AGN luminosity varies considerably with time because it is
determined by the rate of gas inflow to the nuclear region(s) of the merging galaxies. The fractional contribution of the AGN(s) to the total 1-1000 $\micron$ luminosity
is shown in Fig. \ref{fig:agn_frac}. The contribution is most significant during the time periods shortly after the starbursts induced at first pericentric passage and final
coalescence, when the fractional contribution reaches $\sim 25$ and $\sim 75$ per cent, respectively. Thus, near those times, the AGN emission has a significant effect on the
simulated SEDs (see \citealt{Snyder:2013} for a detailed study). Because \magphys does not include a treatment of AGN emission, it is possible that the AGN emission
can affect the ability of \magphys to obtain a satisfactory fit and infer accurate parameters during the time periods of the simulation when the AGN contribution is significant.
Thus, it is worthwhile to check how the results differ when the AGN emission is not included in the radiative transfer calculations.

\begin{figure*}
\centering
\includegraphics[trim=0 377 531 11,clip,width={0.50\columnwidth}]{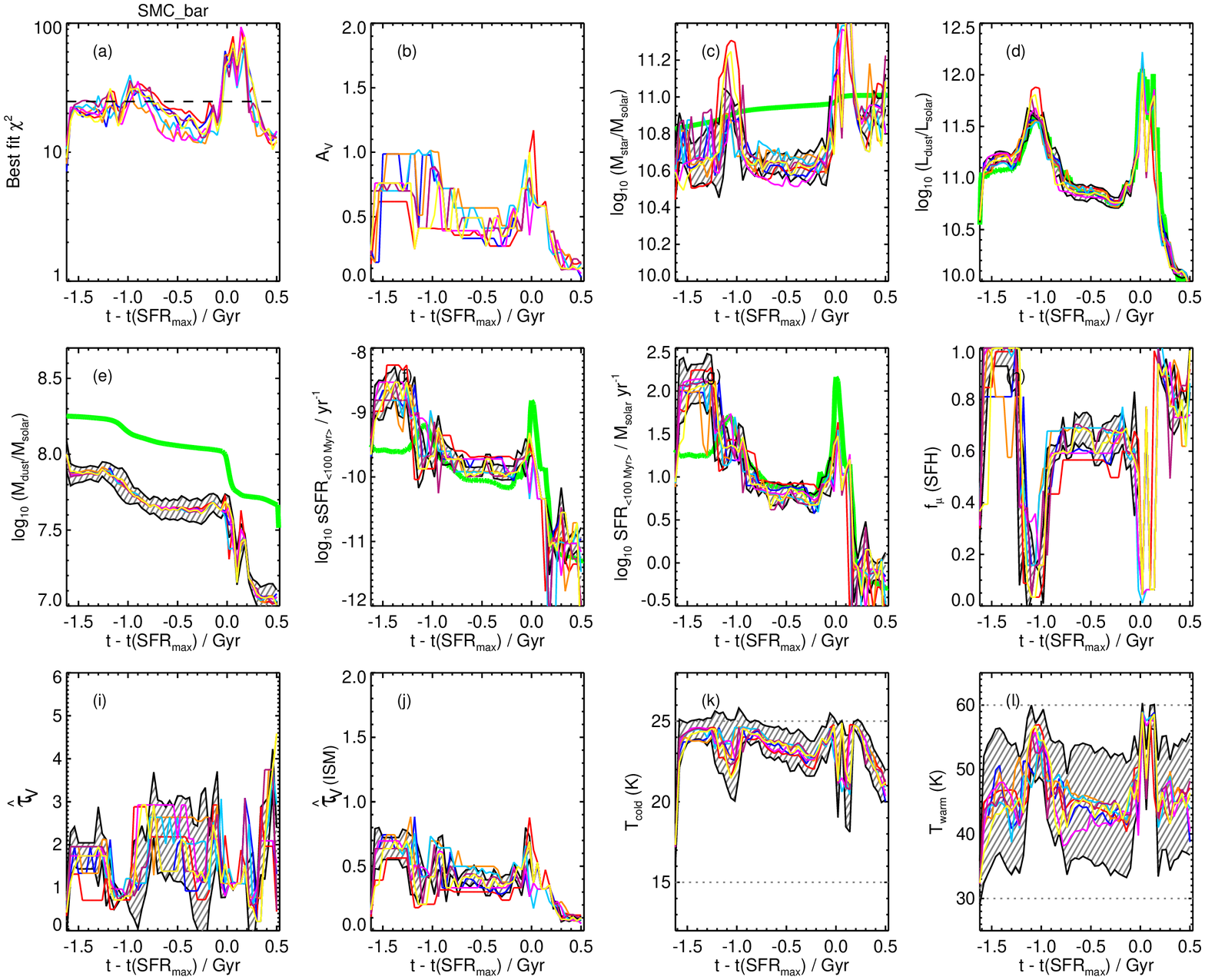}
\includegraphics[trim=354 377 177 0,clip,width={0.50\columnwidth}]{./plots/fitresults_smc_bar.eps}
\includegraphics[trim=354 188 177 189,clip,width={0.50\columnwidth}]{./plots/fitresults_smc_bar.eps}
\includegraphics[trim=531 188 0 189,clip,width={0.50\columnwidth}]{./plots/fitresults_smc_bar.eps}
\caption{Selected results for the \smc test. When SMC-type dust is used, the quality of the \magphys fits decreases considerably, as indicated by the systematically
greater $\chi^2$ values (leftmost panel) compared with the \fiducial case. For much of the evolution of the merger, the fits are only marginally acceptable or unacceptable.
The median-likelihood stellar mass values (second panel from left) agree much less well with the true values than for the \fiducial run; although the fits are formally acceptable
for $-0.8 \la \tmer \la -0.1$ Gyr, \magphys underestimates the stellar mass by as much as $\sim 0.3$ dex. The inferred SFR (third panel from left) can be severely incorrect: for
$-1.6 \la \tmer \la -1.2$ Gyr, \magphys infers an SFR of zero for most viewing angles when the true SFR is $\sim 20 \msunperyr$, which is likely because \magphys
attributes all of the FIR emission to the diffuse ISM rather than the stellar birth clouds (i.e. $\fmusfh = 1$, as indicated in the rightmost panel).}
\label{fig:smc}
\end{figure*}

For the merger simulation, we performed a \sunrise run in which we artificially set the AGN luminosity to zero (the \agnoff run); by comparing the results of this run with the \fiducial
run, we can determine how AGN contamination affects the \magphys results.\footnote{\label{footnote:agn_off}Because we used the same \gadgetthree simulation, which includes black hole
accretion and thermal AGN feedback, this \sunrise calculation is technically not physically self-consistent. However, the virtue of this test is that it enables us to quantify the impact
of the AGN emission on the simulated galaxy SEDs with all else (e.g. the SFH and galaxy geometry, which would be altered had we disabled black hole accretion and AGN feedback
in the \gadgetthree simulation) being equal. See \citealt{Snyder:2013} for a detailed analysis of similar tests.} For most parameters, the \magphys results for the \fiducial and
\agnoff runs do not differ significantly. However, the differences in the $\chi^2$ values and recovered stellar masses are of interest. The time evolution of these two quantities for
the \agnoff run are shown in Fig. \ref{fig:agn_off}. The $\chi^2$ values at the peak of the starburst and AGN activity [$-0.1 \la \tmer \la 0.2$ Gyr] are less when the
AGN emission is disabled, which indicates that AGN contamination is part of the reason that \magphys did not yield satisfactory fits for the \fiducial run during that
phase of the merger (however, the fits are still formally unacceptable at this time for the \agnoff run). For the \agnoff run, the stellar mass is recovered more accurately
than for the \fiducial case (compare panel \pmstarn of Fig. \ref{fig:merger} and the bottom panel of Fig. \ref{fig:agn_off}).
Thus, AGN contamination partially causes the overestimate of the stellar mass in the \fiducial run during the coalescence phase of the merger.
Still, it is reassuring that although \magphys does not account for AGN emission, most of the recovered parameters are robust to AGN contamination.
Even when the AGN contributes $\sim 25$ per cent of the UV--mm luminosity [at $\tmer \sim -1$ Gyr], \magphys is able to obtain acceptable fits to the photometry
and recover the parameters accurately.\footnote{Note that the lack of MIR photometry may be partially responsible for the
robustness of the results to AGN contamination. However, in a similar test, \citet{Michalowski2014} included mock MIR photometry and also found that the stellar
masses were robust to AGN contamination.}

\subsubsection{Dust grain composition} \label{S:dust}

The dust properties are another source of uncertainty in the SED modelling procedure because the dust composition and grain-size distribution
affect the attenuation curve and shape of the dust SED. Dust in the ISM is a complex topic (see \citealt{Draine:2003araa} for a review): even for the
Milky Way,
Large Magellanic Cloud (LMC), and Small Magellanic Cloud (SMC), it is far from trivial to determine the detailed dust properties in some region of the ISM.
Furthermore, the dust properties are likely very different in different regions of the ISM of a galaxy; for example, it is possible that typical grain sizes are greater
in higher-density regions \citep[e.g.][]{Kelly:2012}. Naturally, dust in high-redshift galaxies is even less understood than for local galaxies, and it may be possible
that dust properties vary significantly with redshift because of e.g. the differences in the timescales for the various dust-production channels
\citep[e.g.][]{Valiante:2009,Michalowski:2010production}. Indeed, there is some observational evidence that dust in high-redshift galaxies differs from that in the Milky Way
(e.g. \citealt{Buat:2011,Buat:2012}; \citealt*{Kriek:2013}; \citealt{Aller2014}). Thus, dust is a potentially significant uncertainty inherent in SED modelling that cannot be ignored.

Because we typically do not have a detailed understanding of a galaxy's dust properties when fitting its SED, an empirically supported attenuation curve is typically assumed;
at best, one can use a flexible attenuation curve parameterisation, as is done in \magphys, or multiple attenuation curves to help characterise the significance of this uncertainty.
We have investigated this uncertainty by varying the
intrinsic properties of the dust, which affect the effective attenuation curve and the FIR SED shape, used in the \sunrise calculation. In addition to the default
MW model, we performed \sunrise runs in which the \citet{Draine:2007kk} LMC and SMC models were used. Because the attenuation curve used by
\magphys was not changed, these tests mimic the situation in which the dust properties assumed when fitting a galaxy's SED do not correspond to the
true dust properties of the galaxy.

\begin{figure*}
\centering
\includegraphics[width=1.95\columnwidth,trim=0cm 0cm 0cm 0.65cm,clip]{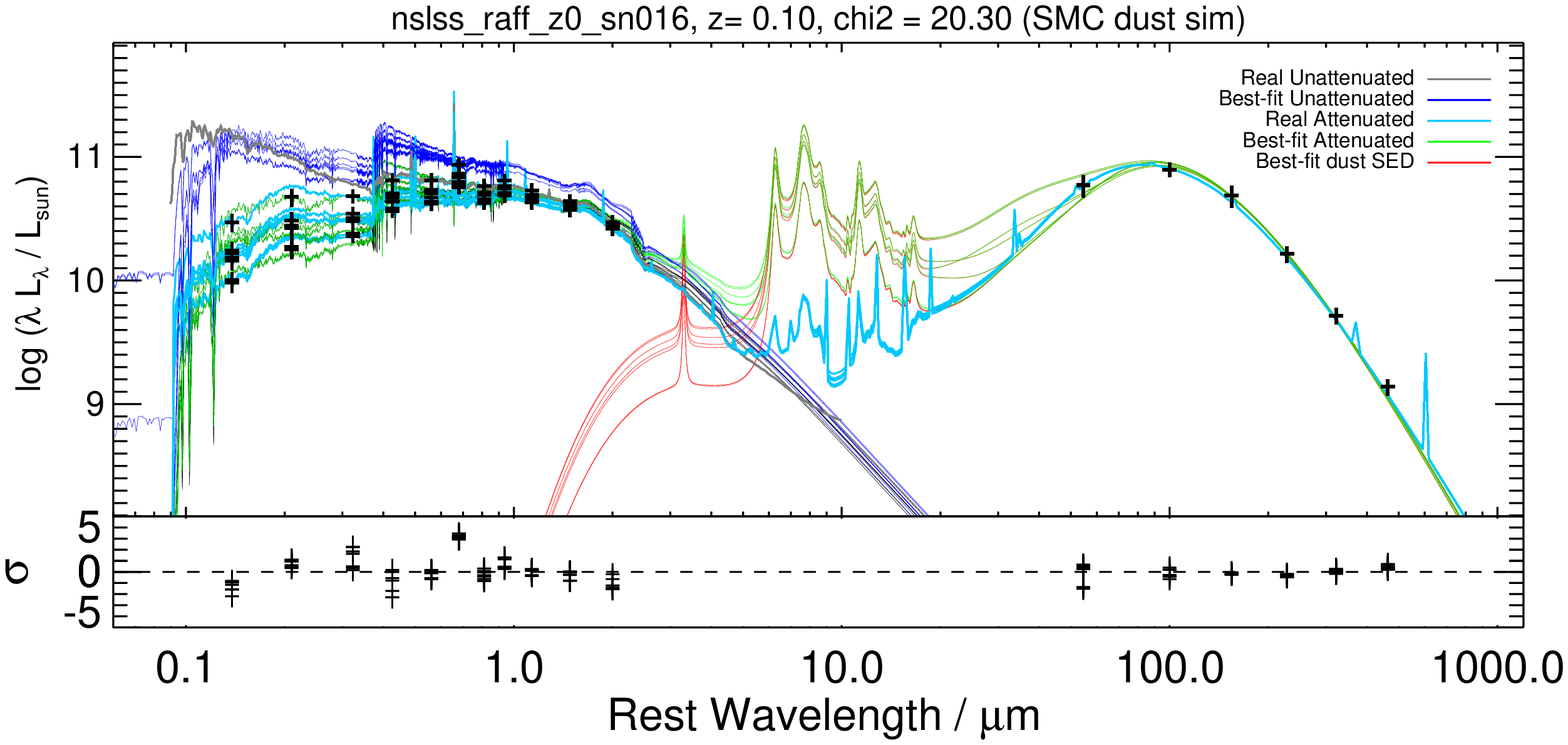}
\caption{Example SED fits for the $\tmer = -1.47$ Gyr snapshot of the \smc run; see the
caption of Fig. \ref{fig:sed_example} for a complete description of what is plotted. Unlike
for the \fiducial case, the intrinsic SEDs inferred by \magphys (blue lines) differ significantly
from the true intrinsic SED (grey line), even though \magphys yields acceptable fits
to the synthetic photometry. Consequently, for some parameters (e.g. the SFR),
\magphys recovers the true values poorly.}
\label{fig:smc_sed}
\end{figure*}

Fig. \ref{fig:lmc} shows selected results for the \lmc run.
The results for when LMC-type dust was used in the radiative transfer calculations
are similar to those of the \fiducial case, for which the MW dust model
was used. The $\chi^2$ values (top panel)
for the \lmc run tend to be greater than for the
corresponding snapshot of the \fiducial run, but the fits are still acceptable
except during the near-coalescence phase of the merger.
The only parameter that differs noticeably is the stellar mass: the median-likelihood values yielded by \magphys for the \lmc run
are marginally ($\sim 0.1$ dex) greater than for the \fiducial run.

Fig. \ref{fig:smc} presents selected results for the \smc case, in which SMC-type
dust was used in the radiative transfer calculations rather than the default
MW-type dust. In this case, the $\chi^2$ values (leftmost panel) are significantly
greater than for the \fiducial case, and for most mock SEDs, \magphys
yields fits that are only marginally acceptable or unacceptable.
For almost all snapshots, the median-likelihood values for the stellar mass
(second panel from left) differ from the true values by $\sim 0.2-0.4$ dex, even
when the fits are formally acceptable [e.g. $-0.8 \la \tmer \la -0.2$ Gyr].
The median-likelihood values for the SFR (third panel from left) and sSFR (not shown)
can also differ significantly. Most notably, for most mock SEDs from the time
period $-1.6 \la \tmer \la -1.2$ Gyr, when the fits are still formally acceptable (although
the $\chi^2$ values are often very close to the threshold for an acceptable fit), 
\magphys infers an SFR of zero, but the true
value is $\sim20 \msunperyr$. The reason for this considerable error can be
understood from the rightmost panel of Fig. \ref{fig:smc}, which shows
$\fmusfh$, the fraction of stellar luminosity that is absorbed by the diffuse ISM
rather than the birth clouds. When the inferred SFR is zero, $\fmusfh = 1$, which
indicates that \magphys has attributed the considerable FIR luminosity
($\lir > 10^{11} \msun$; see panel \pldustn of Fig. \ref{fig:merger}) exclusively
to older stellar populations. 

To understand the origin of this discrepancy, it is
instructive to investigate how well \magphys is able to recover the true intrinsic
SEDs. Fig. \ref{fig:smc_sed} shows an example of the SED fits for the $\tmer = -1.47$
Gyr snapshot of the \smc case. This figure is similar to Fig.
\ref{fig:sed_example}, which shows SED fits for the \fiducial case (refer to the
caption of Fig. \ref{fig:sed_example} for full details regarding what is plotted). For all viewing angles,
\magphys yields acceptable fits to the synthetic photometry. However, the intrinsic
SEDs inferred by \magphys (blue lines) differ considerably from the true intrinsic
SED (grey line). \magphys tends to underestimate (overestimate) the intrinsic UV
(optical through NIR) emission. Because the true intrinsic SED is not recovered
well by \magphys (because the attenuation curve inferred by \magphys differs significantly from the true
attenuation curve; see below), it is unsurprising that the parameter values
yielded by \magphys can differ significantly from the true values. We suggest that this may be less likely to occur for
real observations of actual galaxies than it is in our simulations because the inevitable addition of photometric measurement
errors should ensure larger $\chi^2$ values, thereby making these discrepant fits unacceptable. Indeed, it may also be possible
to `tune' the arbitrary photometric errors assumed in the fitting to alleviate this potential issue, but we make no attempt to do so here.

\begin{figure}
\centering
\includegraphics[width=0.99\columnwidth,trim=0cm 0cm 0cm 0.77cm,clip]{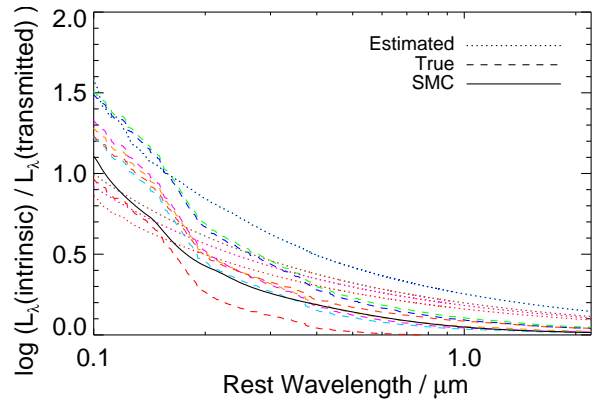}
\caption{Comparison of the true attenuation curves (dashed lines) and attenuation curves
inferred by \magphys (dotted lines) for the $\tmer = -1.47$ Gyr snapshot of the \smc run. The different
colours correspond to different viewing angles. The black solid line denotes the intrinsic SMC-type
opacity curve (with arbitrarily normalisation) that is used in the radiative transfer calculations.
Generically, the true attenuation curves are significantly steeper than those inferred by \magphys.
Consequently, for the \smc case, \magphys is unable to effectively correct for dust, and the
unattenuated SED inferred by \magphys differs considerably from the true intrinsic SED, as
shown in Fig. \ref{fig:smc_sed}. Thus, the recovered values of parameters such as the SFR can 
differ significantly from the true values.}
\label{fig:smc_attenuation_curve}
\end{figure}

Fig. \ref{fig:smc_attenuation_curve} shows a comparison of the true attenuation
curve and the attenuation curve inferred by \magphys for each viewing angle
for the $\tmer = -1.47$ Gyr snapshot of the \smc run (for which the SEDs are shown in Fig. \ref{fig:smc_sed});
the intrinsic SMC dust opacity curve (which has been arbitrarily normalised) is also plotted for comparison.
For all snapshots, the true attenuation curve is significantly steeper than that inferred by \magphys.
Consequently, even if the $\av$ value recovered by \magphys is accurate, \magphys will under-correct
(over-correct) for dust attenuation in the UV (optical through NIR). This effect explains why the
intrinsic SED tends to be underestimated (overestimated) in the UV (optical through NIR), as
shown in Fig. \ref{fig:smc_sed} and described above.

Recall that the shape of the attenuation curve in \magphys depends on the assumptions
of the CF00 model, in which the optical depth scales as $\lambda^{-1.3}$ in the `birth clouds'
and $\lambda^{-0.7}$ in the `diffuse ISM'. In the simulations, the attenuation curve that results for
a given snapshot and viewing angle depends not only on the intrinsic opacity curve of the dust
but also the spatial distribution of stars and dust, which results in differential attenuation, and spatial
variations in age and metallicity, which cause the intrinsic emission to spatially vary. Thus,
it is perhaps unsurprising that the attenuation curves of the simulated galaxies are sometimes not
described effectively by the standard CF00 model.\footnote{However, it may be possible to better
correct for attenuation using the CF00 model by allowing the power-law indices of
$\hat{\tau}_{\lambda}^{\rm{BC}}$ and $\tauvism$ to vary and marginalizing over this additional uncertainty.}

The results for the \smc and, to a lesser extent, \lmc cases highlight the difficulty of accurately
correcting for dust attenuation. Unfortunately, our understanding of the dust grain composition
is still limited, even for relatively nearby galaxies \citep{Amanullah2014,Patat2014}, and there is
evidence that galaxy attenuation curves can systematically vary with galaxy properties
\citep[e.g.][]{Buat:2012,Kriek:2013}. Thus, dust is likely to remain a significant uncertainty in SED modelling for
some time, and one should interpret results that depend sensitively on the assumed attenuation
curve with caution.

\subsubsection{The presence of very cold dust} \label{S:mp_off}

\begin{figure}
\centering
\includegraphics[trim=0 188 531 189,clip,width={0.75\columnwidth}]{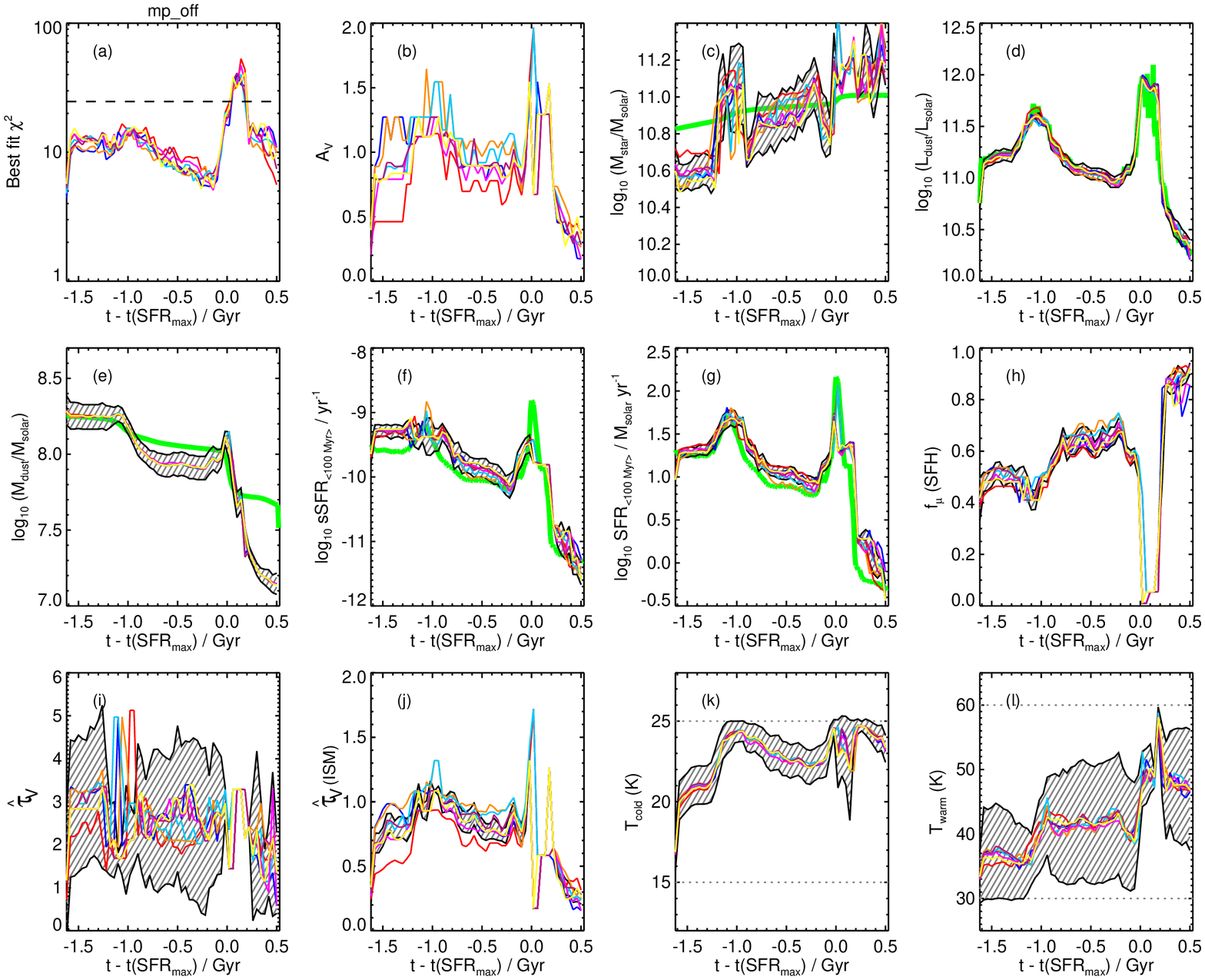}\\
\includegraphics[trim=0 0 531 378,clip,width={0.75\columnwidth}]{./plots/fitresults_mp_off.eps}
\caption{Selected results for the the \altism test, in which all dust in the simulated ISM (rather than just that in the diffuse ISM)
was used. The dust mass (top) is recovered significantly more accurately than for the \fiducial case, although it is still underestimated by
$\la 0.2$ (as much as $\sim 0.6$) dex during the phase between first pericentric passage and final coalescence (post-starburst phase).
Furthermore, the total optical depth (bottom) and $\av$ values (not shown) are greater than in the \fiducial case, which reflects
the fact that the attenuation along any line of sight is guaranteed to be greater in this case than when the default ISM model is used.}
\label{fig:mp_off}
\end{figure}

The \citet{Springel:2003} sub-resolution model implicitly splits the gas contained in a given SPH particle into cold, dense clouds (which contain the bulk of the mass
but have a relatively low volume filling fraction) and a diffuse phase. In the default ISM treatment in \sunrise, it is assumed that the cold phase has negligible volume
filling fraction. Thus, the dust associated with the cold phase does not absorb photons and consequently does not emit radiation. This sub-resolution model
is used as the default model in \sunrise because the real ISM of galaxies is certainly not smooth on $\sim 100$-pc scales. Unfortunately, the resolution of our simulations and
sub-resolution ISM model used prevent us from resolving the full phase structure of the ISM. Simply ignoring the dust in the cold phase of the ISM is a crude
approach for treating this unresolved clumpiness. However, for the purpose of testing how well \magphys can recover the dust mass, it is clearly undesirable
to include dust that does not absorb or emit radiation by construction.

To investigate the uncertainty in the results that is associated with unresolved clumpiness in the ISM, it is also possible to use the dust associated with
both phases of the sub-resolution ISM, which may be more appropriate in some regimes; see \citet{Jonsson:2010sunrise}, \citet{Hayward:2011smg_selection},
\citet{Snyder:2013}, and \citet{Lanz2014} for detailed discussions of this issue. We refer to this alternate treatment of sub-resolution clumpiness of the ISM as
the `alternate ISM' (or `multiphase-off' in the parlance of \citealt{Hayward:2011smg_selection}) model.
When the alternate ISM model is used, each cell in the \sunrise grid contains the same or a greater mass of dust as in the \fiducial case; for regions of high
gas density (e.g. the central starburst), the difference can be an order of magnitude. Consequently, when the alternate ISM model is used, the attenuation along
any line of sight is greater, and the dust temperatures tend to be colder because of dust self-absorption.

Fig. \ref{fig:mp_off} shows the results for the merger simulation when we use the alternate ISM model in the \sunrise calculation, which we refer to as the
\altism case. The trends for most \magphys parameters are qualitatively the same as for the \fiducial case. However, there a few interesting differences.
Most importantly, the dust mass (top panel) is recovered significantly more accurately than for the \fiducial run shown in Fig. \ref{fig:merger}. The reason
for the superior agreement is that when the alternate ISM model is used in the \sunrise calculation, all of the dust in the simulated galaxies can potentially
absorb and emit radiation. In the \fiducial run, the dust in the sub-resolution cold clouds does not absorb or reemit any radiation but is still counted as part of
the total dust mass.

However, even in the \altism case, the dust mass inferred by \magphys is slightly ($\la 0.2$ dex) underestimated during the phase between
first pericentric passage, and it is significantly underestimated (by up to $\sim 0.6$\,dex) during the post-starburst phase. The likely
reason for the remaining underestimate is that after the strong starburst and AGN activity, much of the remaining gas (and thus dust) is contained in a low-density, extended
hot halo. Consequently, the optical depth through this halo is very low, and the dust contained in it absorbs little radiation. As a result, a significant fraction of the dust
cannot be detected via emission. Furthermore, the total optical depth (bottom panel) and $\av$ values (not shown) are typically greater than
in the \fiducial case. This result demonstrates that \magphys qualitatively captures the key physical difference between the two runs.

Unfortunately, the need to use a sub-resolution ISM model in the \sunrise calculations precludes us from determining which case is more correct.
Ideally, use of the next generation of galaxy simulations, which are able to achieve parsec-scale resolution
\citep[e.g.][]{Hopkins:2013mergers,Hopkins:2013FIRE,Hopkins:2013merger_winds}, may eliminate this uncertainty.

\section{Discussion} \label{S:discussion}

\subsection{Dependence on viewing angle}

One of the strengths of our approach is its ability to test how, for a given simulated galaxy, the results of SED modelling vary with viewing angle. Ideally,
estimates of intrinsic physical parameters of a galaxy, such as the SFR and stellar mass, should be insensitive to the perspective from which the galaxy
is observed. For almost all \magphys parameters plotted in this work, for a given simulation snapshot, the median-likelihood values vary with viewing angle
by less than the uncertainty; thus to all intents and purposes, viewing-angle effects do not cause systematic errors in the results.
This result is naturally quite reassuring because for real galaxies, only one viewing angle is available.

Some parameters (primarily $\tauvism$) do vary with viewing angle, but, in so far as the parameters can be interpreted physically, they should depend
on viewing angle. Thus, this variation is not a cause for concern.

\subsection{Other potential sources of error}

In this section, we will briefly discuss other potential sources of error in SED modelling. This issue will be investigated in greater detail in future work.

\subsubsection{Photometric uncertainties}

Throughout this work, no noise was added when generating the mock photometry. Thus, the tests represent the ideal situation in which there are no observational
uncertainties and the inherent physical uncertainties (i.e. those that originate from discrepancies between the model assumptions and reality) are the only source
of discrepancies between the inferred and true parameters (i.e. they are the sole contributors to the best-fit $\chi^2$). These tests are useful for understanding the
fundamental limitations of the method that cannot be
overcome through the use of more-accurate photometry, but they are clearly unrepresentative of the real-world process of modelling galaxy SEDs. Consequently,
it is worthwhile to examine the effects of including observational uncertainties when generating the mock photometry.

We performed a series of tests in which we added a simple Gaussian noise model to the mock photometry for the \fiducial run, and used \magphys to fit the noisy photometry with the same assumed errors discussed in section \ref{S:sims} (similar tests were performed in \citealt{Smith:2012} to validate the consistency of \magphys by feeding it photometry derived from several of the best-fitting SEDs with
simulated Gaussian measurement errors superposed). As expected, the $\chi^2$ values
were greater than for the noiseless case, and \magphys did not yield a statistically acceptable fit for a significantly greater number of mock SEDs.
However, the recovered median-likelihood values for the physical parameters did not differ qualitatively (although the confidence intervals became wider),
and the qualitative evolution of the various physical parameters of the simulation was captured just as well as for the \fiducial case. This result suggests that
the median-likelihood parameter values yielded are robust to the inclusion of realistic random uncertainties and demonstrates the effectiveness of the
Bayesian fitting method employed by \magphys.

\subsubsection{SED coverage}

The results of SED modelling can potentially depend on the wavelength sampling of the photometry used \citep[e.g.][]{Smith:2012,Pforr:2012,Pforr:2013}.
In this work, the photometric bands used are those that were available for the initial \textit{H}-ATLAS investigations, which provide relatively
good coverage of the SEDs in the UV--NIR and FIR; MIR data are noticeably absent. Including MIR data could potentially change the results
significantly. For example, MIR data may help to better constrain the relative contributions of young and old stellar populations to the dust heating.
However, the MIR tends to be sensitive to the presence of AGN (see e.g. \citealt{Snyder:2013} for a detailed discussion). Thus, inclusion of MIR
data could make it significantly more difficult to fit the synthetic SEDs using \magphys.

Because galaxy surveys vary considerably in
terms of the available photometry, it would be worthwhile to investigate the effects of varying the photometry used in the SED modelling.
As a first test, we investigated the effects of excluding the PACS photometry. Although the agreement was generally good in the comparatively
quiescent phases of the simulation, the most significant discrepancy was that the IR luminosity and SFR
were underestimated by $\sim 0.5$ dex during the starburst that occurs at first passage (but it is possible that this underestimate could be
corrected by modifying the priors; E. da Cunha, private communication). This
test further highlights the importance of the available photometry sampling the peak of the temperature-dependent SED for the purposes of recovering the true dust luminosity,
in agreement with the investigation by \citet{Smith:2013} which used isothermal models to fit the dust SEDs of {\it H}-ATLAS galaxies.

\subsubsection{Emission lines}

Another potential source of uncertainty is the contribution of nebular emission lines (which are typically not accounted for by SED modelling
codes) to the broadband photometry \citep[e.g.][]{Charlot:2001,Schaerer:2009,Pacifici:2012,Schenker:2013,Stark:2013}.
At certain redshifts, especially
$z \sim 6-7$, not accounting for contamination from nebular emission can cause the stellar ages \citep{Schaerer:2009} and stellar masses
\citep{Schenker:2013,Stark:2013} to be overestimated. Our simulated SEDs include nebular emission lines; thus, they can contribute
to the broadband photometry. Indeed, the contribution of H$\alpha$ emission is the cause for the larger residuals near the $i$ band that can be
observed in Fig. \ref{fig:sed_example}, and this effect is also often seen in the SED fits of \hatlas\ galaxies in \citet{Smith:2012}. \magphys
is able to consider H$\alpha$ emission as part of the input data set (although these data were unavailable at the time that \citealt{Smith:2012} was written);
we defer a detailed investigation of the influence of emission lines on the derived SED parameters to a future investigation.

\subsection{Applicability of the results to other SED modelling codes}

It is important to keep in mind that we have only employed one SED modelling code, \magphys, which has multiple advantages,
including the following: 1. it utilizes
the full UV--mm SED, and including information yielded by the dust emission can potentially break degeneracies that could not be addressed
using UV--NIR data alone. 2. The underlying SFHs are continuous SFHs with superimposed random bursts. Consequently, it is not subject
to the potential systematic errors that are associated with single-component SFHs \citep[e.g.][]{Michalowski:2012,Michalowski2014}.
3. Through its use of the CF00 dust attenuation model, differential attenuation of young stellar populations can be
(approximately) accounted for.

Because \magphys represents a relatively sophisticated, state-of-the-art SED modelling code, its success at recovering the physical properties
of our simulated galaxies cannot be generalised to all SED modelling codes. Thus, it would be worthwhile to perform similar tests for other
commonly used SED modelling codes. As a first step,
\citet{Michalowski2014} tested the ability of multiple SED modelling codes to recover the stellar masses of simulated submm galaxies
(SMGs). They found that as long as a single-component SFH was not used, all of the codes were able to accurately recover the stellar masses, albeit with a
factor of $\sim 2$ uncertainty. However, this work was deliberately limited in scope to the stellar masses of SMGs, and a more comprehensive
comparison of SED modelling codes is warranted.

\subsection{Recommendations for applying SED modelling codes}

We have demonstrated that for the \fiducial runs, \magphys recovered the true physical parameter values of the simulated galaxies well.
However, uncertainties in the `microphysics', especially regarding the dust attenuation law, can cause serious discrepancies between
the median-likelihood parameter values output by \magphys and the true values \emph{even when the fits are formally acceptable}
(although as we have discussed in section \ref{S:dust}, this should only affect a small fraction of SED fits).
Consequently, for real galaxies, for which e.g. the dust attenuation law or IMF may vary with galaxy properties, there is a risk of making
significant errors for some subset of the observed galaxy population when attempting to recover the physical parameters of the galaxies
through SED modelling.

Because SED modelling is now applied to datasets that contain hundreds to hundreds of thousands of galaxies, it is infeasible to check
the individual fits one-by-one to search for irregularities. One approach for avoiding significant mis-estimates of physical parameters
would be to use a significantly more conservative $\chi^2$ threshold than what was used in this work. However, this would result in
discarding many galaxies for which the vast majority of fits are acceptable and the parameters are well recovered, which is clearly undesirable.

Perhaps the best approach is to broaden the experimental priors in an attempt to `marginalise' over our ignorance. This could be achieved, for
example, by comparing the results derived using multiple distinct SED modelling
approaches; ideally, the approaches should utilise different assumptions about e.g. the dust attenuation (see e.g. \citealt{Bolzonella:2000},
\citealt{Burgarella:2005}, and \citealt{Buat:2011,Buat:2012} for examples). Furthermore, simpler techniques, such
as using empirical laws to estimate the SFR from $\lir$ or radio continuum luminosity, should also be used; although these certainly have their own
caveats, they can still provide additional insight, and current `panchromatic' SED fitting codes lack the machinery to include radio continuum data
in their analyses. For objects for which the results of
different SED modelling approaches or/and simpler techniques differ, one should interpret the results with caution and investigate further.
Such disagreements are especially likely for galaxies that differ significantly from the galaxies that were used for validation of the model,
as is the case for \magphys (with the standard priors) and submm galaxies \citep{Rowlands2014}.

Such a multi-faceted validation may seem tedious, and it would naturally require more human effort and computational time. However, we
believe that the additional investment will be rewarded with significantly more robust results, or, at the least, a determination of the
types of galaxies for which (some of) the physical properties must remain `known unknowns' for the time being.

\section{Conclusions} \label{S:conclusions}

By applying the SED modelling code \magphys to synthetic photometry generated by performing dust radiative transfer on hydrodynamical simulations of
an isolated disc galaxy and a galaxy merger, we have investigated how well \magphys can recover the intrinsic properties of the simulated galaxies. Our
principal conclusions are the following:
\begin{enumerate}
\item For the isolated disc galaxy simulation, \magphys yields acceptable fits at all times. The $V$-band attenuation, stellar mass, dust luminosity, SFR, and sSFR
are recovered accurately. The dust mass is systematically underestimated, but whether this underestimation will occur for real galaxies is unclear (see conclusion vii).
\item For the galaxy merger simulation, when the assumptions regarding the IMF, SSP models, and dust composition in \magphys and the dust radiative transfer
calculations are similar, \magphys yields acceptable fits and recovers all parameters except the dust mass well, except during the near-coalescence phase of
the merger, when the starburst and AGN activity are most intense. During this phase, the fits are often not formally acceptable, but most parameters are
still recovered reasonably well.
\item For most parameters, the variation in the median-likelihood values with
viewing angle is less than the uncertainty for a single viewing angle. For
parameters that should depend on viewing angle, such as $\tauvism$, the variation
with viewing angle can be greater than the uncertainty for a single viewing angle.
\item Although \magphys does not include AGN emission, the galaxy properties that we infer are generally unaffected by AGN contamination.
Even when the AGN contributes as much as 25 per cent of the UV--mm luminosity, \magphys can obtain statistically acceptable fits to the photometry
and recover the parameters accurately.
\item When either LMC- or SMC-type (rather than the default MW-type) dust are used to perform radiative transfer to calculate the mock photometry,
\magphys recovers some parameters
less well. For the \lmc case, the median-likelihood stellar mass values are $\sim 0.1$ dex greater but still consistent with the true values within the uncertainties. When
SMC-type dust is used, \magphys yields marginally acceptable or unacceptable fits for the majority of the mock SEDs. Most
notably, for some snapshots for which the SFR is $\sim 20 \msunperyr$, \magphys yields median-likelihood SFR values of zero even though the fits
are formally acceptable.
\item The amount by which the dust mass is underestimated depends on the sub-resolution ISM model used in the radiative transfer calculations. In the best-case
scenario, \magphys recovers the dust mass well during the first-passage and coalescence phases of the merger but underestimates it by $\sim 0.1-0.2$ dex (as
much as $\sim 0.6$ dex) during the phase between first passage and coalescence (post-starburst phase).
\end{enumerate}

Overall, our results constitute a somewhat mixed endorsement of the SED modelling approach: when the assumptions made regarding e.g. the dust attenuation
curve are relatively consistent with the true attenuation curve, \magphys performs very well. However, if, for example, the true dust attenuation curve
differs significantly from that assumed by \magphys, one may be better served by using less-sophisticated but more transparent methods for inferring
physical properties of galaxies from their SEDs. Regardless, one should use caution when performing SED modelling on large samples of galaxies
and ideally cross-check the results by using multiple SED modelling codes and comparing with the results of simpler techniques.

\acknowledgments

We are very grateful to the anonymous referee, Elisabete da Cunha, Matt Jarvis and Kate Rowlands for detailed comments on the manuscript.
We thank Lauranne Lanz, Micha{\l} Micha{\l}owski and R\"udiger Pakmor for useful discussions
and Volker Springel for providing the non-public version of \gadgetthree used for this work.
CCH acknowledges the hospitality of the Aspen Center for Physics, which is supported by the National Science Foundation Grant No. PHY-1066293, and
the Centre for Astrophysics at the University of Hertfordshire, and he is grateful to the Klaus Tschira Foundation and Gordon and Betty Moore Foundation for financial support.
DJBS acknowledges the hospitality of the Heidelberg Institute for Theoretical Studies.
This research has made use of NASA's Astrophysics Data System Bibliographic Services.
\\

\footnotesize{
\bibliography{std_citations,smg,sed,additional_citations}
}

\begin{appendix}

\section{Full fitting results for alternate-assumption \sunrise runs}

For completeness, in this appendix, we present the full fitting results for the \sunrise runs in which the physical assumptions
were varied. The most interesting panels of these figures were already presented and discussed above.

\begin{figure*}
\centering
\includegraphics[width=2.0\columnwidth,trim=0cm 0cm 0cm 0.35cm,clip]{./plots/fitresults_agn_off.eps}
\caption{Similar to Fig. \ref{fig:merger}, but for the \agnoff run.}
\label{fig:agn_off_full}
\end{figure*}

\begin{figure*}
\centering
\includegraphics[width=2.0\columnwidth,trim=0cm 0cm 0cm 0.35cm,clip]{./plots/fitresults_lmcavg_10.eps}
\caption{Similar to Fig. \ref{fig:merger}, but for the \lmc run.}
\label{fig:LMC_full}
\end{figure*}

\begin{figure*}
\centering
\includegraphics[width=2.0\columnwidth,trim=0cm 0cm 0cm 0.35cm,clip]{./plots/fitresults_smc_bar.eps}
\caption{Similar to Fig. \ref{fig:merger}, but for the \smc run.}
\label{fig:SMC_full}
\end{figure*}

\begin{figure*}
\centering
\includegraphics[width=2.0\columnwidth,trim=0cm 0cm 0cm 0.35cm,clip]{./plots/fitresults_mp_off.eps}
\caption{Similar to Fig. \ref{fig:merger}, but for the \altism run.}
\label{fig:mp_off_full}
\end{figure*}

\end{appendix}

\label{lastpage}

\end{document}